\documentclass[12pt,a4paper]{article}
\usepackage{amsmath,caption,amsfonts,setspace,mathrsfs,mathtools,amsthm}
\usepackage[top=1.18in, bottom=1.18in, left=1.1in, right=1.1in]{geometry}

\usepackage[round]{natbib}
\usepackage{color,soul}

\DeclareCaptionStyle{italic}[justification=centering]{labelfont={bf},textfont={it},labelsep=colon}
\usepackage{graphicx,psfrag,epsf,textcomp}
\usepackage{enumerate, dsfont}
\usepackage{natbib}
\usepackage{url,xcolor} 
\usepackage{booktabs, subfig, bm, paralist,mathpazo,tikz,todonotes,longtable,microtype}
\usepackage[pdftex,colorlinks=true]{hyperref}
\definecolor{darkblue}{rgb}{0,0,.6}
\hypersetup{citecolor=darkblue,linkcolor=darkblue,urlcolor=darkblue}

\date{}

\newcommand{\blind}{0}

\graphicspath{{plots/}}

\newsavebox\CBox

\newcommand{\bbP}{{\bf P}}

\begin{document}

\def\spacingset#1{\renewcommand{\baselinestretch}{#1}\small\normalsize} \spacingset{1}

\if0\blind
{
  \title{\bf Double bootstrapping for visualising the distribution of descriptive statistics of functional data}
  \author{
    Han Lin Shang\footnote{Corresponding address: Department of Actuarial Studies and Business Analytics, Level 7, 4 Eastern Road, Macquarie University, Sydney, NSW 2109, Australia; Email: hanlin.shang@mq.edu.au; ORCID: \url{https://orcid.org/0000-0003-1769-6430}} \\
Department of Actuarial Studies and Business Analytics \\
Macquarie University
 }
  \maketitle
} \fi

\if1\blind
{
    \title{\bf Double bootstrapping for visualising the distribution of descriptive statistics of functional data}
  \maketitle
} \fi

\vspace{-.45in}

\begin{abstract}
We propose a double bootstrap procedure for reducing coverage error in the confidence intervals of descriptive statistics for independent and identically distributed functional data. Through a series of Monte Carlo simulations, we compare the finite sample performance of single and double bootstrap procedures for estimating the distribution of descriptive statistics for independent and identically distributed functional data. At the cost of longer computational time, the double bootstrap with the same bootstrap method reduces confidence level error and provides improved coverage accuracy than the single bootstrap. Illustrated by a Canadian weather station data set, the double bootstrap procedure presents a tool for visualising the distribution of the descriptive statistics for the functional data.

\vspace{9pt}
\par
\end{abstract}

\def\thefigure{\arabic{figure}}
\def\thetable{\arabic{table}}

\renewcommand{\theequation}{\thesection.\arabic{equation}}

\fontsize{12}{14pt plus.8pt minus .6pt}\selectfont

\setcounter{equation}{0}
\def\spacingset#1{\renewcommand{\baselinestretch}{#1}\small\normalsize} \spacingset{1}
\spacingset{1.5}

\section{Introduction}

Recent computer technology in data collection and storage allows statisticians to analyse functional data. The monographs by \cite{RS02, RS05} present state-of-art parametric techniques, while the book by \cite{FV06} gives a nonparametric treatment for functional data analysis. For selective reviews on various aspects of functional data analysis, consult \cite{Cuevas14}, \cite{Shang14}, \cite{Morris15}, \cite{GV16}, \cite{WCM16} and \cite{RGS+16}. 

In functional data analysis, a key objective is to draw statistical inference about the distribution of descriptive statistics from realisations generated from an unknown data generating process at a population level. Although it is important to have a consistent estimator for descriptive statistics, it is equally important to estimate the variability associated with the descriptive statistics, construct confidence intervals (CIs) and carry out hypothesis tests. When such a problem arises, bootstrapping turns out to be the only practical alternative \citep[see, e.g.,][]{CFF06, MP11}.

Since functional data do not have a well-defined density, nonparametric bootstrapping is commonly used for independent and identically distributed (i.i.d.) functional data. In nonparametric bootstrapping, one resamples $\bm{\mathcal{X}}^b= \{\mathcal{X}_1^b,\dots,\mathcal{X}_n^b\}$ drawn randomly with replacement from original functions $\bm{\mathcal{X}}=\{\mathcal{X}_1,\dots,\mathcal{X}_n\}$, so that each $\mathcal{X}_i^b$ has probability $n^{-1}$ of being equal to any given one of the $\mathcal{X}_j$'s, 
\begin{equation*}
\bbP\left(\mathcal{X}_i^b = \mathcal{X}_j|\bm{\mathcal{X}}\right)=n^{-1},\qquad 1\leq i, j\leq n.
\end{equation*}
With a set of bootstrap samples $\{\bm{\mathcal{X}}^{1}, \cdots, \bm{\mathcal{X}}^{B}\}$, we can reduce coverage level error between the nominal and empirical coverage probabilities by iterating the bootstrap once more, where $B$ denotes the number of bootstrap samples. Through a nonparametric bootstrap, one resamples $\bm{\mathcal{X}}^{b\eta} = \{\mathcal{X}_1^{b\eta},\dots,\mathcal{X}_n^{b\eta}\}$ drawn randomly with replacement from bootstrap functions $\bm{\mathcal{X}}^{b}=\{\mathcal{X}_1^{b},\dots,\mathcal{X}_n^{b}\}$, so that each $\mathcal{X}_i^{b\eta}$ has probability $n^{-1}$ of being equal to any given one of the $\mathcal{X}_j^b$'s,
\begin{equation*}
\bbP\big(\mathcal{X}_i^{b\eta} = \mathcal{X}_j^{b}|\bm{\mathcal{X}}^b, \bm{\mathcal{X}}\big) = n^{-1},\qquad 1\leq i,j\leq n.
\end{equation*}

We use the single and double bootstrap procedures to study the distribution of descriptive statistics of functional data denoted by $\theta=\psi(\mathcal{F}_0)$, which is a function of the underlying distribution function $\mathcal{F}_0$. The bootstrap principle is to estimate a function of the population distribution $\mathcal{F}_0$ by using the same function of the bootstrap distribution function based on a random sample of size $n$ from $\mathcal{F}_0$. Following the early work by \cite{Martin90b, Martin90}, we consider an example of iterated bootstrapping, namely double bootstrapping, to construct CIs with more accurate coverage probability than single bootstrapping for estimating the distribution of descriptive statistics of functional data \citep[see also][]{Shang15}. 

There is an extensive literature on double bootstrapping for scalar-valued data \citep[see, e.g.,][]{CH15}. The first article to mention the double bootstrap was by \cite{Hall86}, followed quickly by \cite{Beran87, Beran88}. Extensive theoretical discussions are given in \cite{Hall88} and \cite{HM88}. In the statistics literature, double bootstrapping was used to construct CIs and estimate the underlying distribution \citep[see, e.g.,][]{BH94, BP98, LY99, CH15}. The double bootstrapping was also incorporated within the errors-in-variables framework and applied to a calibration problem \citep[see][]{Pevsta13}. In the econometrics literature, \cite{DM02} uses the double bootstrapping to improve the reliability of bootstrap tests of non-nested linear regression models. The contribution of this paper is to extend the use of double bootstrapping to functional data analysis. Since functional data are often observed with measurement error, we aim to provide accurate uncertainty quantification for the descriptive statistics of functional data.

The outline of this article is as follows: Section~\ref{sec:2} sets notations, definitions, and introduces a bootstrap method. Section~\ref{sec:3} presents the descriptive statistics, the $L_2$ distance metrics used for evaluation and comparison, and the finite sample performance based on a series of simulation studies. In Section~\ref{sec:4}, we apply the single and double bootstrap procedures to a Canadian weather station data set. Section~\ref{sec:6} concludes, along with some ideas on how the double bootstrap procedure presented here can be further extended. Additional simulation results are shown in the appendices.

\section{Bootstrapping}\label{sec:2}

\subsection{Notation}

Random functions are assumed to be sampled from a second-order stochastic process $\mathcal{X}$ in $\mathcal{L}^2[0,\tau]$, where $\mathcal{L}^2[0,\tau]$ is the Hilbert space of square-integrable functions on the bounded interval $[0,\tau]$. The stochastic process $\mathcal{X}$ satisfies the finite variance condition $\int^{\tau}_0\mathcal{X}^2(t)dt<\infty$, inner product $\langle f, g \rangle = \int^{\tau}_{0}f(t)g(t)dt$ for any two functions, $f$ and $g \in \mathcal{L}^2[0,\tau]$ and induced squared norm $\|\cdot\|^2 = \langle \cdot, \cdot \rangle$. All random functions are defined on a common probability space $(\Omega, A, P)$. The notation $\mathcal{X} \in \mathcal{L}_{\mathcal{H}}^2(\Omega, A, P)$ is used to indicate $\text{E}(\|\mathcal{X}\|^p)<\infty$ for some $p>0$, where $\mathcal{H}$ denotes the Hilbert space. When $p=1$, $\mathcal{X}(t)$ has the mean curve $\mu(t) = \text{E}[\mathcal{X}(t)]$; when $p=2$, a non-negative definite covariance function is given by
\begin{equation}
c_{\mathcal{X}}(t,s) = \text{Cov}[\mathcal{X}(s),\mathcal{X}(t)] = \text{E}\left\{[\mathcal{X}(s)-\mu(s)][\mathcal{X}(t) - \mu(t)]\right\}, \label{eq:1}
\end{equation}
for all $s,t\in \mathcal{I}$.

The covariance function $c_{\mathcal{X}}(t,s)$ in~\eqref{eq:1} allows the covariance operator of $\mathcal{X}$, denoted by $\mathcal{K}_{\mathcal{X}}$, to be defined as
\begin{equation*}
\mathcal{K}_{\mathcal{X}}(\phi)(s) = \int_{\mathcal{I}}c_{\mathcal{X}}(t,s)\phi(t)dt.
\end{equation*}
Via Mercer's lemma, there exists an orthonormal sequence $(\phi_k)$ of continuous function in $\mathcal{L}^2[0,\tau]$ and a non-increasing sequence $\lambda_k$ of positive number, such that
\begin{equation*}
c_{\mathcal{X}}(t,s) = \sum^{\infty}_{k=1}\lambda_k\phi_k(t)\phi_k(s),\qquad s, t\in \mathcal{I},
\end{equation*}
where $(\lambda_1,\lambda_2,\dots)$ are eigenvalues, where $(\phi_1(t),\phi_2(t),\dots)$ are orthogonal eigenfunctions.

\subsection{Bootstrap functions}\label{sec:boot_fun}

Denote $n$ realisations of a stochastic process $\mathcal{X}$ evaluated at $t\in \mathcal{I}$ as $\{\mathcal{X}_1(t),\dots,\mathcal{X}_n(t)\}$. When functional data are i.i.d., we can obtain the bootstrap sample $\{\mathcal{X}_1^b(t),\dots,\mathcal{X}_n^b(t)\}$ directly, by randomly sampling with replacement from the original functions \citep[see also][]{Shang15}. In practice, we can only observe and evaluate $\mathcal{X}$ at discretised data points $0\leq t_1< t_2\cdots <t_T\leq \tau$, thus the discretized bootstrap samples are obtained as $\{\bm{\mathcal{X}}^b(t_j) =[\mathcal{X}^b_1(t_j),\mathcal{X}^b_2(t_j),\dots,\mathcal{X}_n^b(t_j)]^{\top}$; $j = 1,2,\dots,T\}$.

To avoid the possible appearance of repeated curves in the bootstrap samples, \cite{CFF06} and \cite{Shang15} replaced the standard i.i.d. bootstrap samples by so-called smooth bootstrap samples, which are drawn from a smooth empirical distribution function. This can be achieved by adding white noise to the bootstrap sample, expressed as
\begin{equation}
\bm{\mathcal{X}}^0(t_j) = \bm{\mathcal{X}}^b(t_j) + \bm{z}(t_j), \qquad j=1,2,\dots,T,\label{eq:smoothness}
\end{equation}
where $\{\bm{z}(t_j) = [z_1(t_j), z_2(t_j),\dots,z_n(t_j)]^{\top}\}$; and $[\bm{z}(t_1),\bm{z}(t_2),\dots,\bm{z}(t_T)]$ is normally distributed with mean $\bm{0}$ and covariance matrix $\beta\bm{\Sigma}_{\bm{\mathcal{X}}}$; and $\bm{\Sigma}_{\bm{\mathcal{X}}}$ is the covariance matrix of $[\mathcal{X}(t_1),\mathcal{X}(t_2),\dots,\mathcal{X}(t_T)]$ \citep[see also][]{Shang15}. Following \cite{CFF06}, we set $\beta=0.05$. 

\subsection{Confidence level estimation}

Let $\theta := \theta(t) = \psi(\mathcal{F}_0)$ be a statistic whose distribution is unknown, where $\psi$ is the function that defines the parameter $\theta$. Let $c_n(\alpha)$ denote the largest $(1-\alpha)^{\text{th}}$ quantile of the distribution of $\theta$. Suppose $\mathcal{F}$ is the set of all possible values of parameter $\theta$. Then
\begin{equation*}
\{\vartheta\in \mathcal{F}: \theta_{\vartheta}\leq c_n(\alpha)\}
\end{equation*}
is a confidence set of level $1-\alpha$ for $\theta$.

Although the distribution of a statistic is unknown, we can approximate it by an empirical distribution of $\mathcal{F}_1$. Let $\widehat{c}_n(\alpha)$ denote the largest $(1-\alpha)^{\text{th}}$ quantile of the empirical distribution of $\theta$, and the confidence level will typically converge to $1-\alpha$ as the sample size $n$ increases \citep{Beran87}. However, in a finite sample, there is a difference in confidence level error. We aim to study this error via single and double bootstrapping.

Let $\bm{\mathcal{X}}^b$ be a bootstrap sample of size $n$ drawn from the empirical distribution $\mathcal{F}_1$, where each component of $\bm{\mathcal{X}}^b$ is conditionally independent, given the original sample $\bm{\mathcal{X}}$. Denote $\mathcal{F}_1^b$ as the estimate of $\mathcal{F}_0$ re-computed from the bootstrap sample $\bm{\mathcal{X}}^{b}$, and let $\bm{\mathcal{X}}^{b\eta}$ be a double bootstrap sample of size $n$ drawn from the estimated distribution $\mathcal{F}_1^b$, where each component of $\bm{\mathcal{X}}^{b\eta}$ is conditionally independent given $\bm{\mathcal{X}}$ and $\bm{\mathcal{X}}^b$ \citep{Beran87}. 

Let $D(\widehat{\theta},\theta)$ be the distance of a statistic between the population and sample levels. Let $D(\widehat{\theta}^b, \widehat{\theta})$ be the distance of a statistic between the original sample and single bootstrap samples. Let $D(\widehat{\theta}^{b\eta},\widehat{\theta}^b)$ be the distance of a statistic between the single and double bootstrap samples. The empirical coverage probability of $\theta$ obtained from the single bootstrap is given as
\begin{equation*}
\frac{\#\left\{D(\widehat{\theta}^b, \widehat{\theta})>D(\widehat{\theta},\theta)\right\}}{B_1},\qquad b=1,\dots,B_1,
\end{equation*}
where $B_1$ is the number of samples in the single bootstrap. In contrast, the empirical coverage probability of $\theta$ obtained from the double bootstrap is given as
\begin{equation*}
\frac{\#\left\{D(\widehat{\theta}^{b\eta}, \widehat{\theta}^b)>D(\widehat{\theta},\theta)\right\}}{B_1B_2},
\end{equation*}
where $B_2$ is the number of samples in the second layer of the double bootstrap. To reduce both computational time and confidence level error, $B_2$ can equal to one \citep[see, e.g.,][]{CH15}. The difference between the single and double bootstrapping stems from the distance measure $D(\widehat{\theta}^{b\eta},\widehat{\theta})$, which plays the role of error correction to achieve better calibration.

\section{Simulation study}\label{sec:3}

The single and double bootstrap procedures are used to estimate the distribution of descriptive statistics of functional data. In Section~\ref{sec:31}, we review several descriptive statistics that characterise i.i.d. functional data. Section~\ref{sec:32} introduces a simulated Gaussian process, while the evaluation metrics of estimation accuracy are given in Section~\ref{sec:33}. Section~\ref{sec:34} displays the simulation results, where the performance of single and double bootstrapping are evaluated and compared based on their differences between the empirical and nominal coverage probabilities.

\subsection{Descriptive statistics of functional data}\label{sec:31}

We seek an estimator of the functional median, which allows us to rank a set of functions based on their location depth, i.e., the distance from the functional median (the deepest curve). This leads to the idea of functional depth, which has received much attention in the functional data literature \citep[see e.g.,][]{CFF06, CFF07,Gervini12,LR09}. We consider two functional depth measures, namely \citeauthor{FM01}'s \citeyearpar{FM01} depth and \citeauthor{Gervini12}'s \citeyearpar{Gervini12} depth based on small ball probability.

\citeauthor{FM01}'s \citeyearpar{FM01} depth is the oldest functional depth measure. For each $t\in [0,\tau]$, let $\mathcal{F}_{1}$ be the empirical sampling distribution of $\{\mathcal{X}_1(t),\dots,\mathcal{X}_n(t)\}$ and let $Z_i(t)$ be the univariate depth of function $\mathcal{X}_i(t)$, given by
\begin{equation*}
I_i = \int^{\tau}_{0} Z_i(t) dt = \int^{\tau}_{0} 1-\Big|\frac{1}{2} - \mathcal{F}_{1}[\mathcal{X}_i(t)]\Big|dt,
\end{equation*}
and the values of $I_i$ provide a way of ranking the curves from inward to outward \citep{Shang15}. The functional median is the deepest curve with the largest value of $I_i$.

\cite{Gervini12} considered the set of distances between any two functions, denoted by $d(\mathcal{X}_i, \mathcal{X}_j)$. An observation $\mathcal{X}_i$ is an outlier if it is far away from most other functions. Given a probability $\alpha\in [0,1]$, \cite{Gervini12} defines the $\alpha$-radius $r_i$ as the distance between $\mathcal{X}_i(t)$ and its $\lceil\alpha n \rceil^{\text{th}}$ closest observations, where $\lceil x\rceil$ denotes the integer closest to $x$ from above. Customarily, we consider $\alpha = 0.5$. The rank of $r_i$ provides a measure of outlyingness of $\mathcal{X}_i(t)$; the smaller the $r_i$ is, the more dense the $\mathcal{X}_i(t)$ is \citep{Shang15}. The functional median is the curve with the smallest $r_i$. 

Apart from the functional median, we also consider sample versions of the functional mean and functional variance, given by
\begin{align*}
\overline{\mathcal{X}}(t) &= \frac{1}{n}\sum^n_{i=1}\mathcal{X}_i(t),\\
\widehat{V}(t) &= \frac{1}{n-1}\sum^n_{i=1}\left[\mathcal{X}_i(t) - \overline{\mathcal{X}}(t)\right]^2
\end{align*}
and the $\gamma$-trimmed functional mean, which is the mean function of the $100(1-\gamma)\%$ deepest curves. The functional trimmed mean is expressed as
\begin{equation*}
\frac{1}{n-\lceil \gamma n\rceil}\sum^{n-\lceil \gamma n\rceil}_{i=1}\mathcal{X}_i(t),
\end{equation*}
where $\left(\mathcal{X}_{(1)}, \mathcal{X}_{(2)}, \dots, \mathcal{X}_{(n)}\right)$ are the ordered sample curves based on their increasing location depth, and $\gamma\in \Big[0,\frac{n-1}{n}\Big]$ is the trimming parameter. Customarily, we consider $\gamma = 0.05$.

\subsection{Simulation setup}\label{sec:32}

A series of Monte Carlo simulation studies is implemented to evaluate and compare the finite sample performance of the single and double bootstrap procedures for estimating the distribution of descriptive statistics of i.i.d. functional data. For comparison, we consider the functional model previously studied in \cite{CFF06} and \cite{Shang15}: a Gaussian process $\mathcal{X}(t)$ with mean $m(t) = 0.95\times 10t(1-t)+0.05\times 30t(1-t)$, $\text{Cov}[\mathcal{X}(t_i),\mathcal{X}(t_j)]=\exp(-|t_i-t_j|/0.3)$, and $\text{Var}[\mathcal{X}(t)]=1$.

Given that the data generating process is a Gaussian process, we can draw samples from a multivariate normal distribution with mean $[m(t_1), m(t_2)$,$\dots$, $m(t_T)]^{\top}$ and covariance matrix $\text{Cov}[\mathcal{X}(t_i),\mathcal{X}(t_j)] = \min(t_i, t_j) = t_i$. In practice, we evaluate and compare our methods on a common set of grid points. In our simulation studies, we have taken 101 equally-spaced grid points between 0 and 1, for two sample sizes $n=100$ and 300.

\subsection{Simulation evaluation}\label{sec:33}

To evaluate the performance of the bootstrap procedures, we calculate the bootstrap CIs of the descriptive statistics. Given a set of original functions $\{\mathcal{X}_1,\dots, \mathcal{X}_n\}$, we draw $B_1 = B_2=399$ bootstrap samples for each of the $R=200$ replications; and the same pseudo-random seed was used for all the methods to ensure the same simulation randomness \citep[see also][]{Shang15}. For each replication, the $100(1-\delta)\%$ bootstrap CIs of a descriptive statistic $\theta$, are defined by calculating the cut-off value $D(\mathcal{X}_1,\dots,\mathcal{X}_n)$, such that $100(1-\delta)\%$ of the bootstrap repetitions $(\widehat{\theta}^b, b=1,\dots,B_1)$ are within a distance smaller than $D(\mathcal{X}_1,\dots, \mathcal{X}_n)$. In the double bootstrap, the $100(1-\delta)\%$ bootstrap CIs of a descriptive statistic are defined by calculating the cut-off value $D(\mathcal{X}_1^b, \dots, \mathcal{X}_n^b)$, such that $100(1-\delta)\%$ of the bootstrap repetitions $(\widehat{\theta}^{b\eta}, \eta = 1,\dots,B_2)$ are within a distance smaller than $D(\mathcal{X}_1^b,  \dots, \mathcal{X}_n^b)$. 

We calculate the empirical coverage probability that the target function at the population level lies within the CIs. As a performance measure, $L_2$ distance metrics are used for constructing CIs. They are defined as
\begin{align*}
\left\|\theta(t) - \widehat{\theta}(t)\right\|_2 &= \left\{\int_{t\in \mathcal{I}} \left[\theta(t) - \widehat{\theta}(t)\right]^2dt\right\}^{\frac{1}{2}}, \\
\left\|\widehat{\theta}(t) - \widehat{\theta}^b(t)\right\|_2 &= \left\{\int_{t\in \mathcal{I}} \left[\widehat{\theta}(t) - \widehat{\theta}^b(t)\right]^2dt\right\}^{\frac{1}{2}},\\
\left\|\widehat{\theta}^{b}(t) - \widehat{\theta}^{b\eta}(t)\right\|_2 &= \left\{\int_{t\in \mathcal{I}} \left[\widehat{\theta}^{b}(t) - \widehat{\theta}^{b\eta}(t)\right]^2dt\right\}^{\frac{1}{2}}.
\end{align*}
In Appendix A, we present additional simulation results using the $L_{\infty}$ distance metric. 

\subsection{Simulation results}\label{sec:34}

\begin{figure}[!htbp]
\centering
\includegraphics[width=8.4cm]{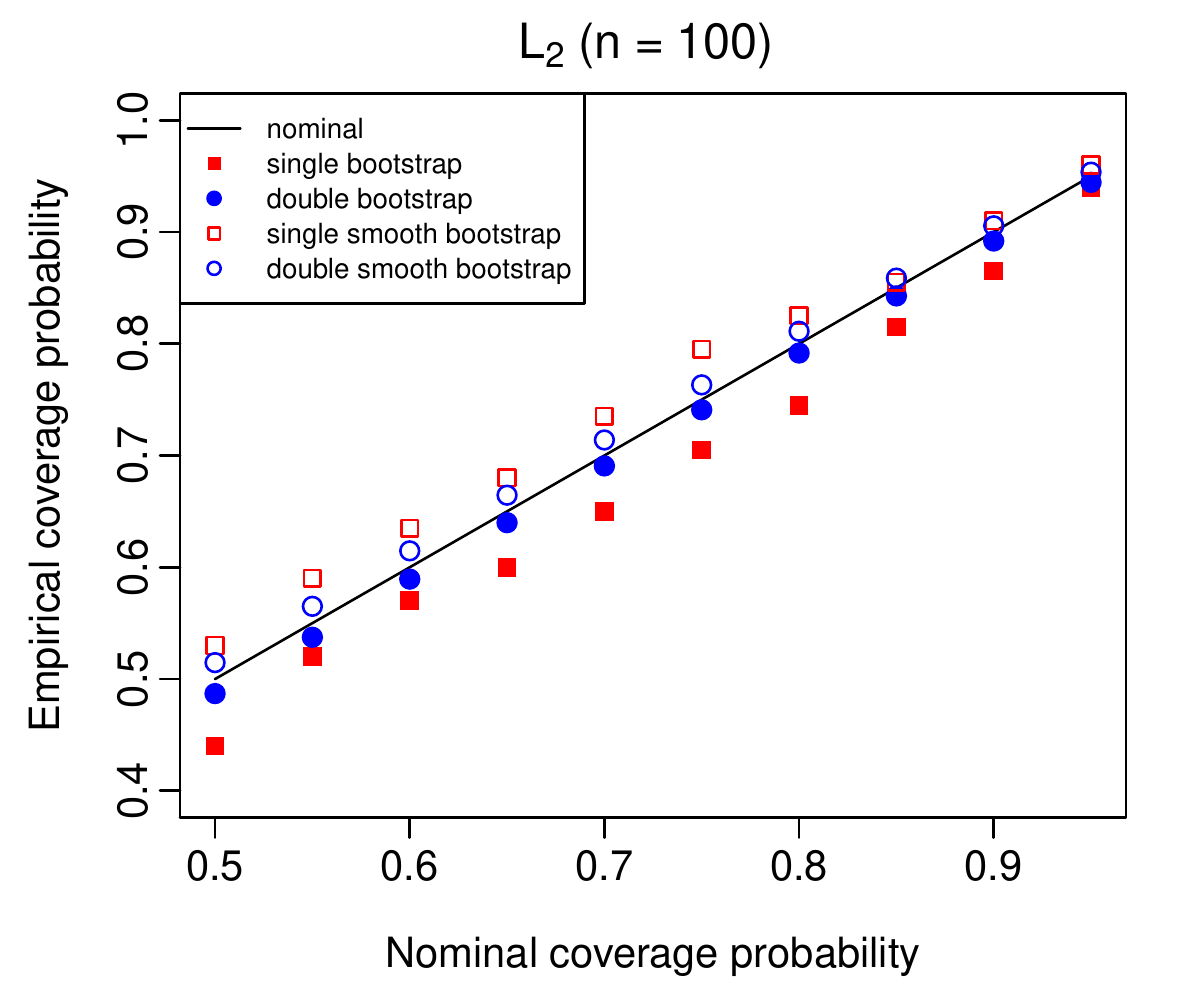}\qquad
\includegraphics[width=8.4cm]{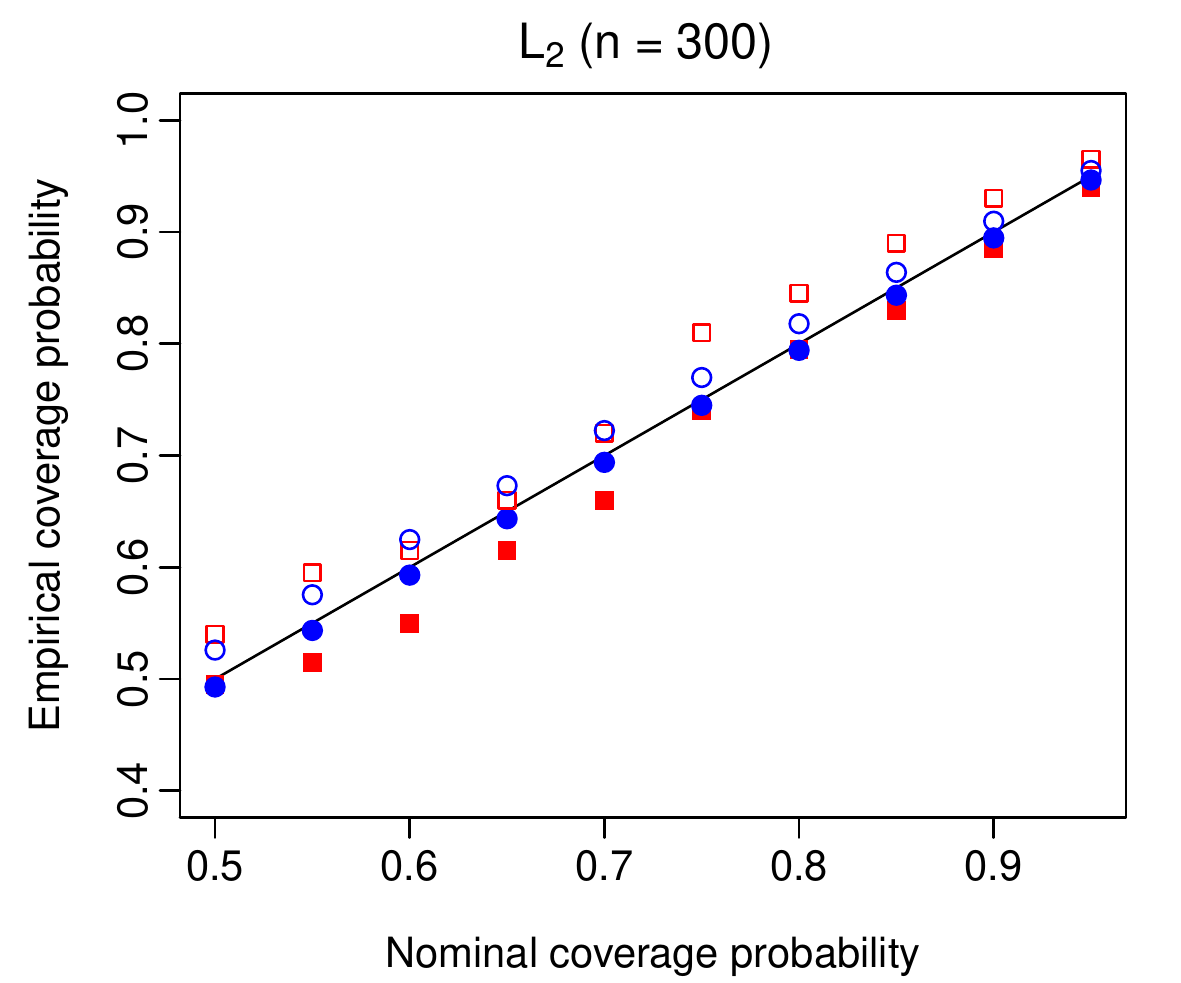}
\caption{Empirical and nominal coverage probabilities for estimating the functional mean, based on $B_1 = B_2 =399$ repetitions and $R=200$ replications. See~\eqref{eq:smoothness} for the smooth bootstrap.}\label{fig:1}
\end{figure}

In Figure~\ref{fig:1}, we plot the nominal (from 0.5 to 0.95 in a step of 0.05) and empirical coverage probabilities for estimating the functional mean of i.i.d. functional data using single and double bootstrap procedures. As the sample size increases from $n=100$ to 300, the empirical coverage probability improves for all the bootstrap procedures, including the single bootstrap procedure of \cite{Shang15}. This result is not surprising, as the validity of bootstrapping relies on a moderate or large sample size \citep[see also][]{MP11}. Subject to the same pseudo-random seed, the double bootstrap procedure outperforms the single bootstrap procedure for most if not all confidence levels. This result is not surprising, as iterating the bootstrap principle reduces the dependence between the probability distribution of the resample and the unknown data generating process \citep[as outlined in][]{Shang15}. Between bootstrapping the original functions and smoothed functions in~\eqref{eq:smoothness}, there is an advantage to bootstrap the original function.

\begin{figure}[!htbp]
\centering
\includegraphics[width=8.4cm]{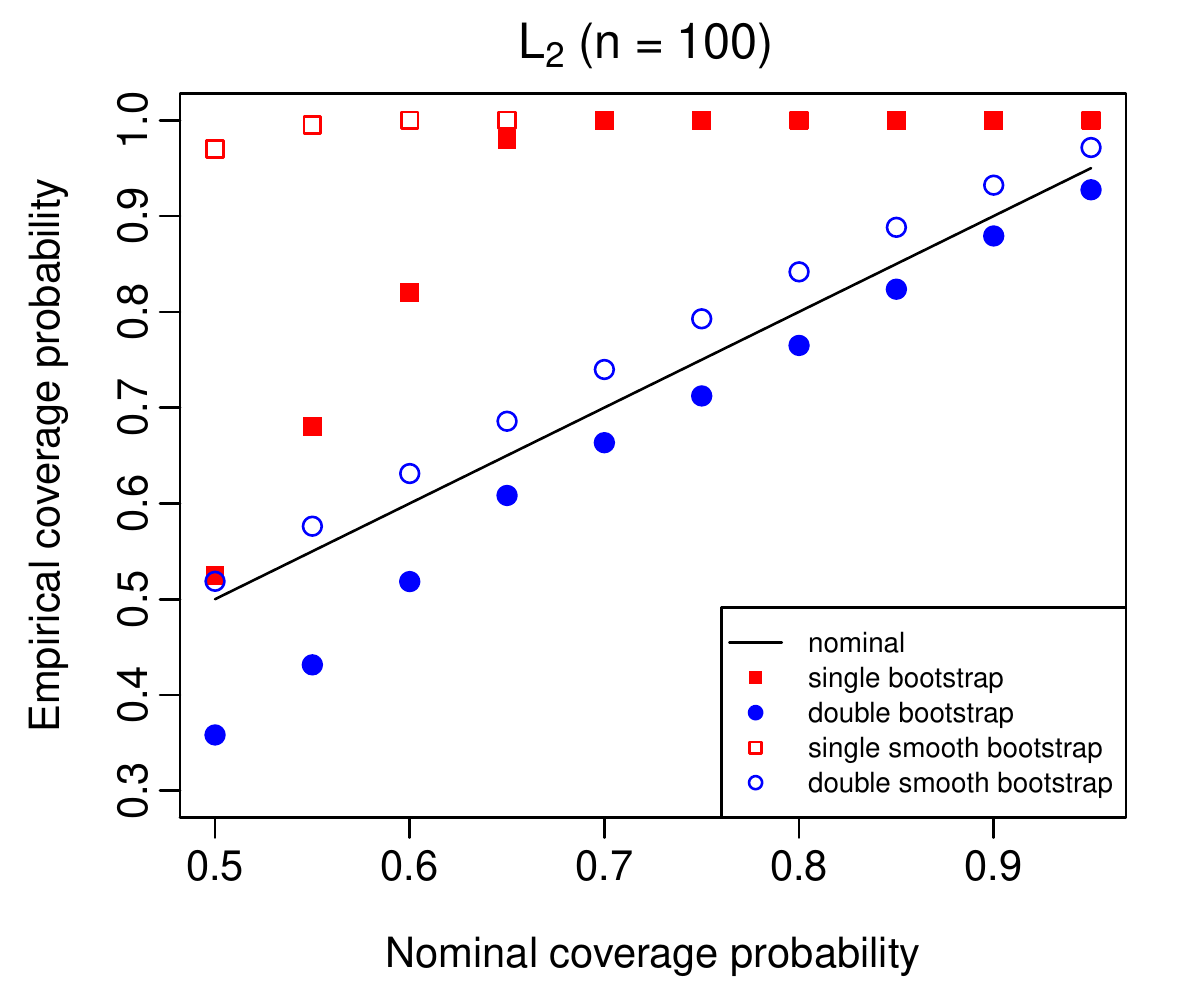}\qquad
\includegraphics[width=8.4cm]{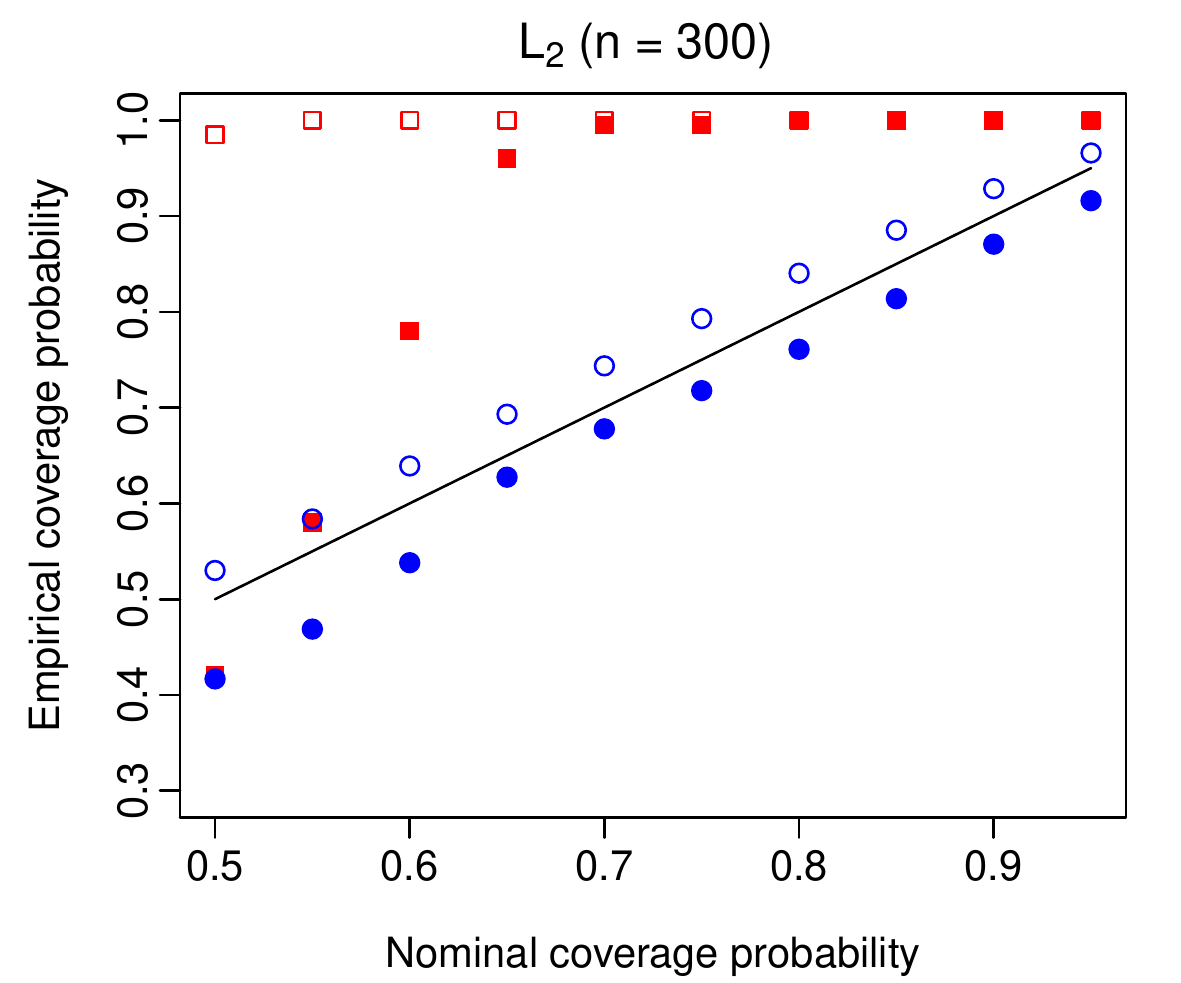}
\caption{Empirical and nominal coverage probabilities for estimating the functional median based on the \citeauthor{FM01}'s \citeyearpar{FM01} depth, using $B_1 = B_2 =399$ repetitions and $R=200$ replications.}\label{fig:2}
\end{figure}

\begin{figure}[!htbp]
\centering
\includegraphics[width=8.5cm]{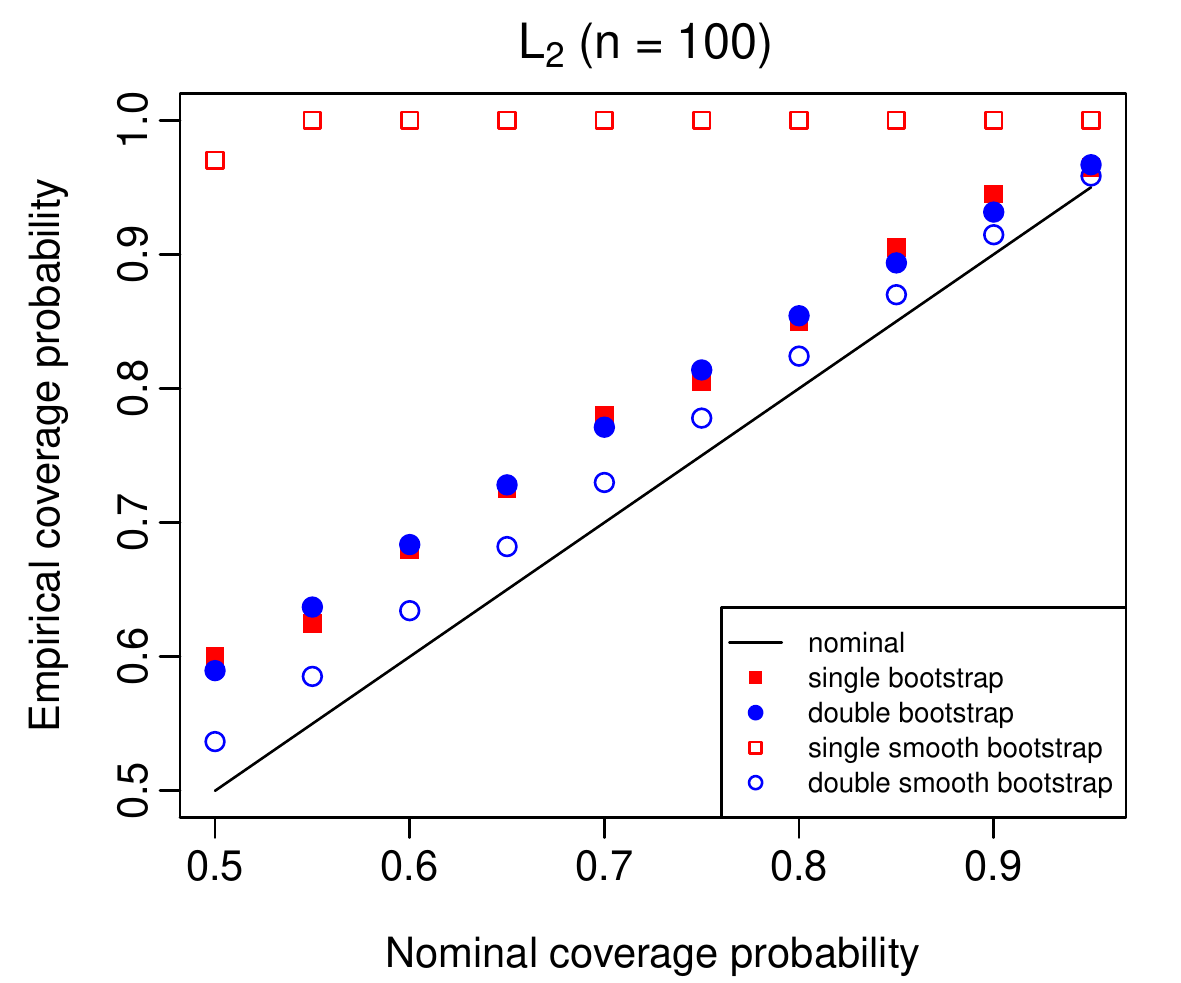}\qquad
\includegraphics[width=8.5cm]{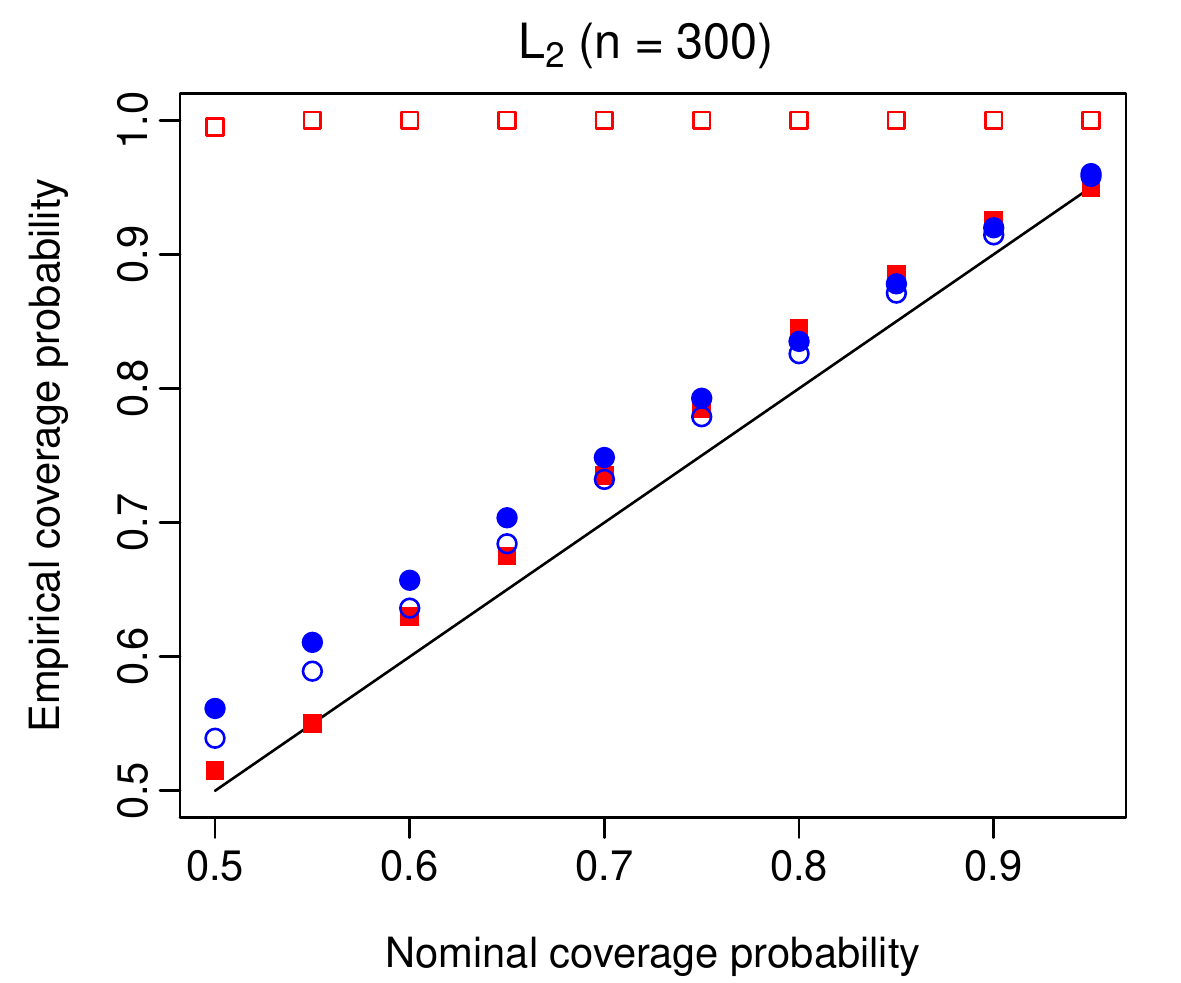}
\caption{Empirical and nominal coverage probabilities for estimating the functional median based on the $\alpha$-radius depth, using $B_1 = B_2 =399$ repetitions and $R=200$ replications.}\label{fig:3}
\end{figure}

In Figures~\ref{fig:2} and~\ref{fig:3}, we present the nominal and empirical coverage probabilities for estimating the functional median based on the \citeauthor{FM01}'s \citeyearpar{FM01} depth and the $\alpha$-radius depth. A striking feature is that the single bootstrap procedure noticeably over-estimates the nominal coverage probability, especially when the confidence level is large. In contrast, the double bootstrap procedure corrects the coverage error, and it even slightly under-estimates the nominal coverage probability. Compared to the functional mean, in general, the functional median is harder to estimate correctly. This is because the quantity being estimated, such as a quantile, is sensitive to the discreteness of empirical distribution function \citep[see, e.g.,][]{BF81, BS85, HM89, ET91}. However, the comparison result further demonstrates the usefulness of the double bootstrap procedure, which can still correct coverage error to achieve better calibration, as measured by the $L_2$ metrics. Between bootstrapping the original and smoothed functions, it is advantageous to bootstrap smoothed function.

\begin{figure}[!htbp]
\centering
\includegraphics[width=8.5cm]{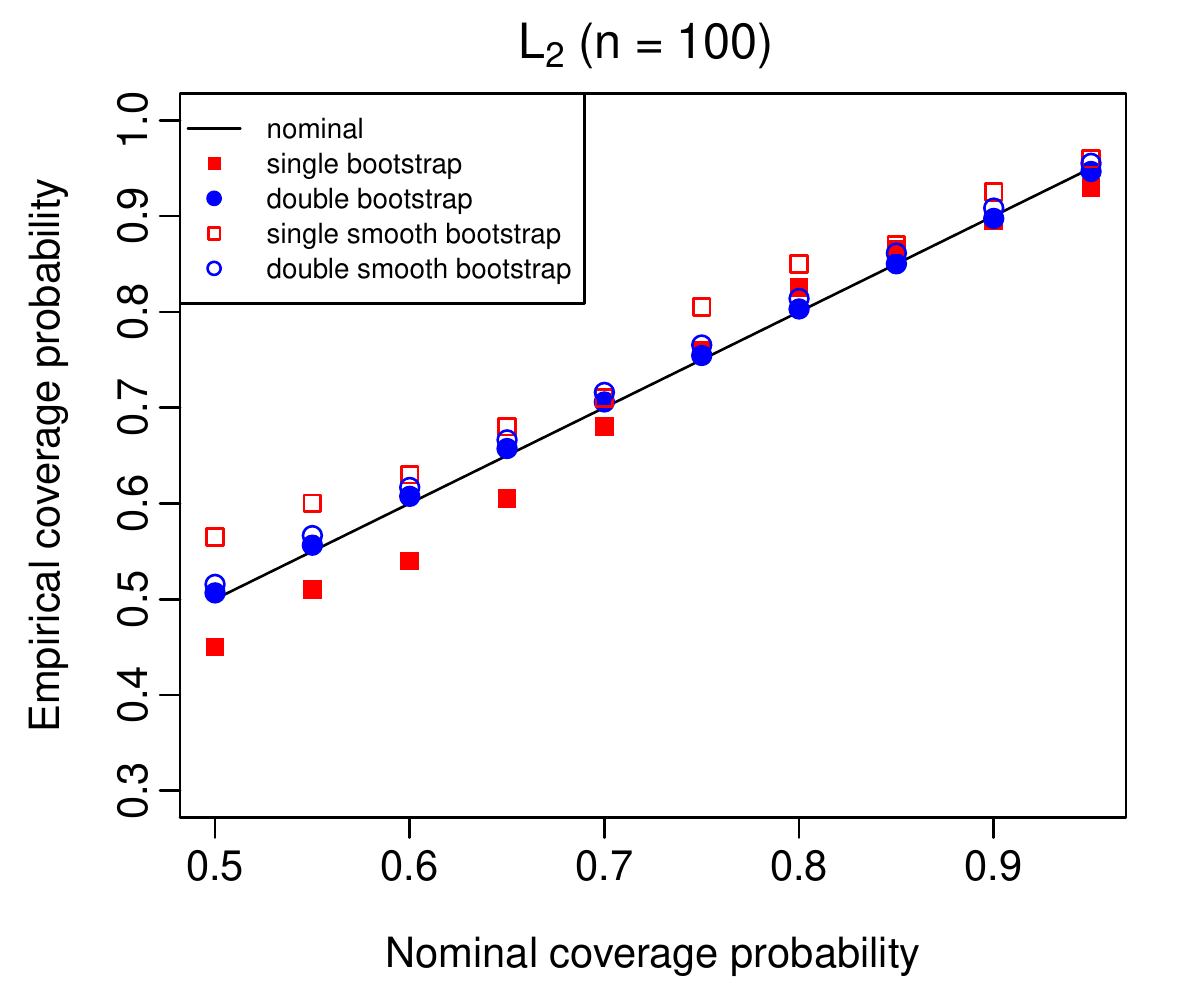}\qquad
\includegraphics[width=8.5cm]{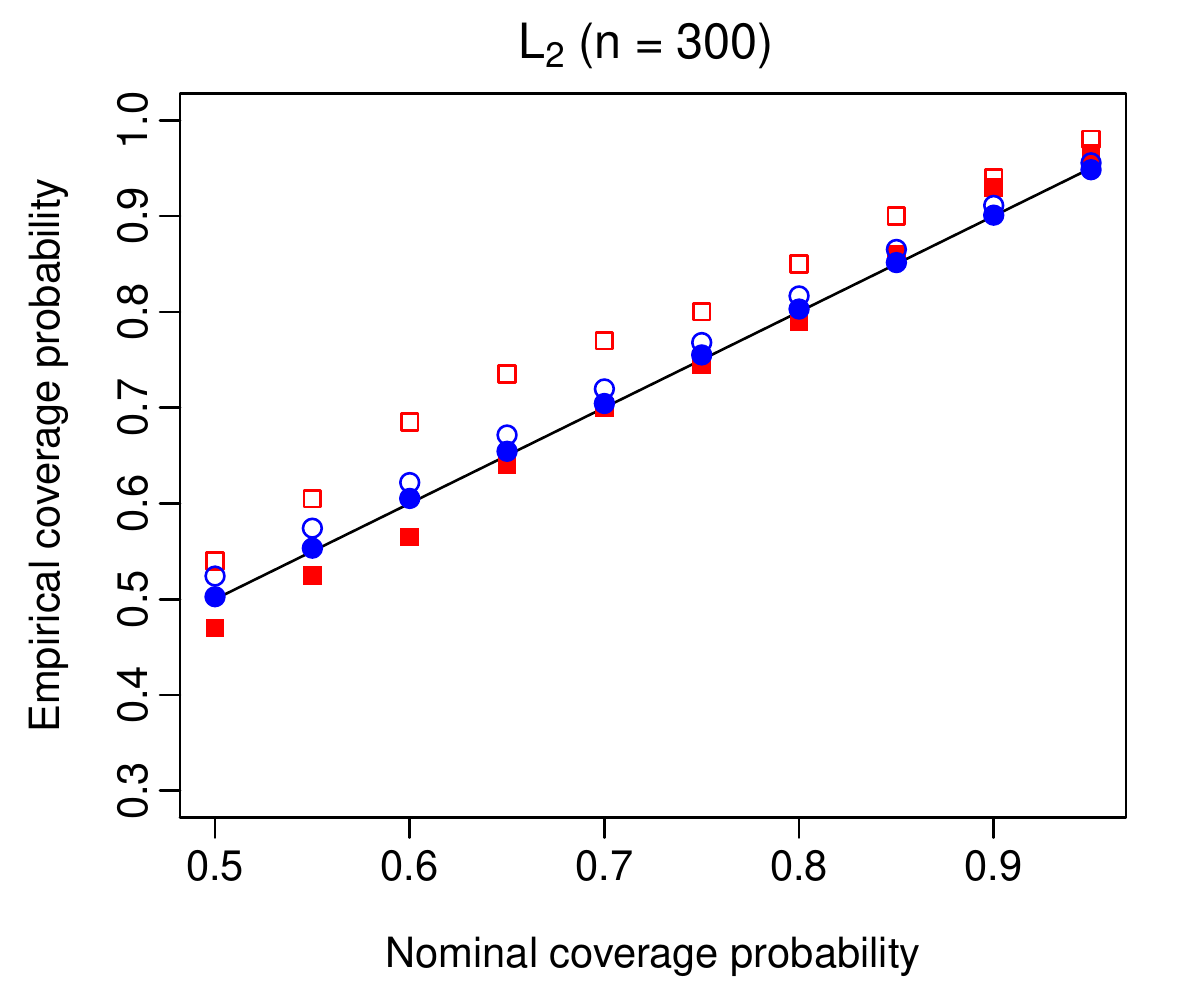}
\caption{Empirical and nominal coverage probabilities for estimating the functional trimmed mean based on the \citeauthor{FM01}'s \citeyearpar{FM01} depth, using $B_1 = B_2 =399$ repetitions and $R=200$ replications.}\label{fig:4}
\end{figure}

\begin{figure}[!htbp]
\centering
\includegraphics[width=8.5cm]{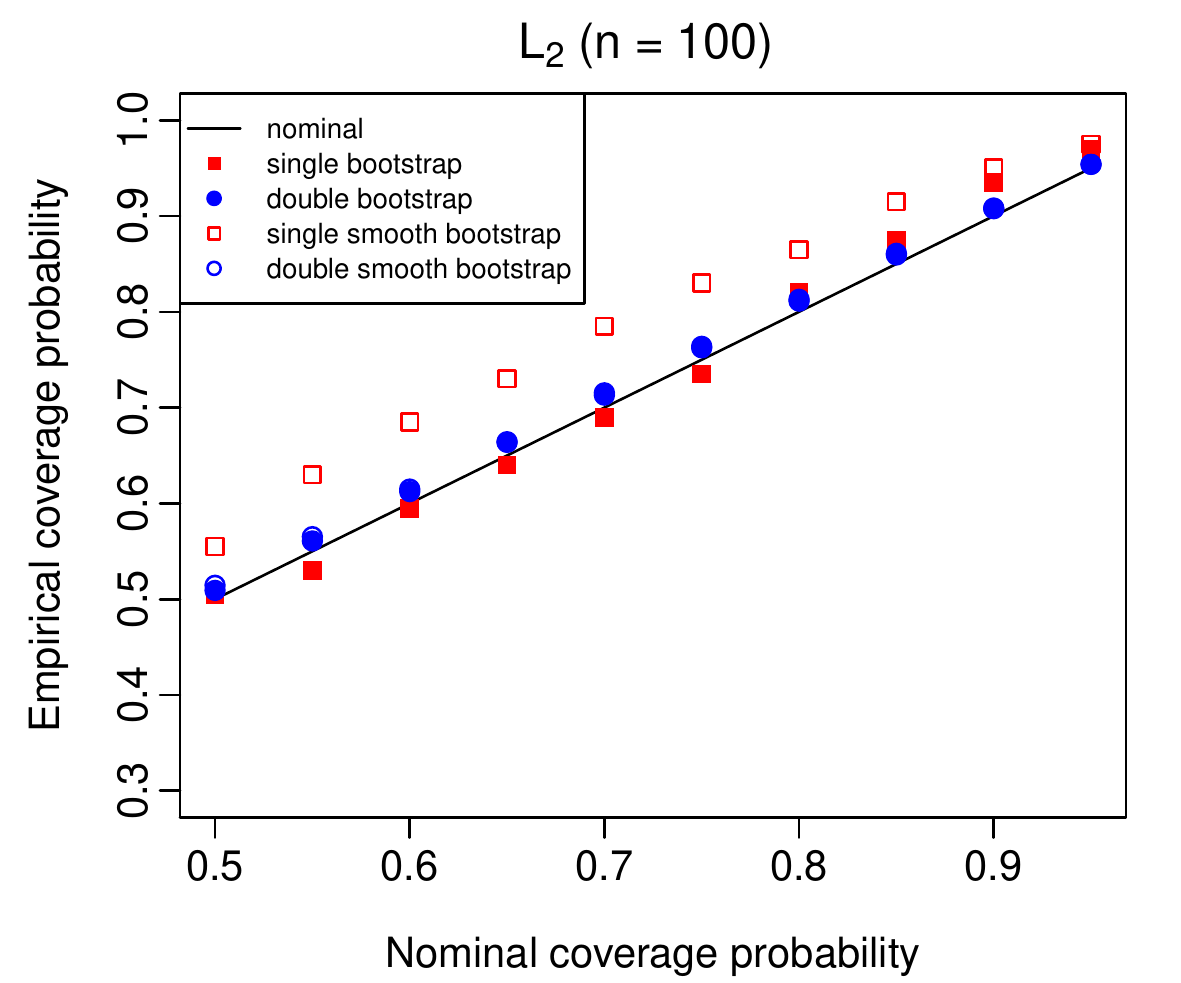}\qquad
\includegraphics[width=8.5cm]{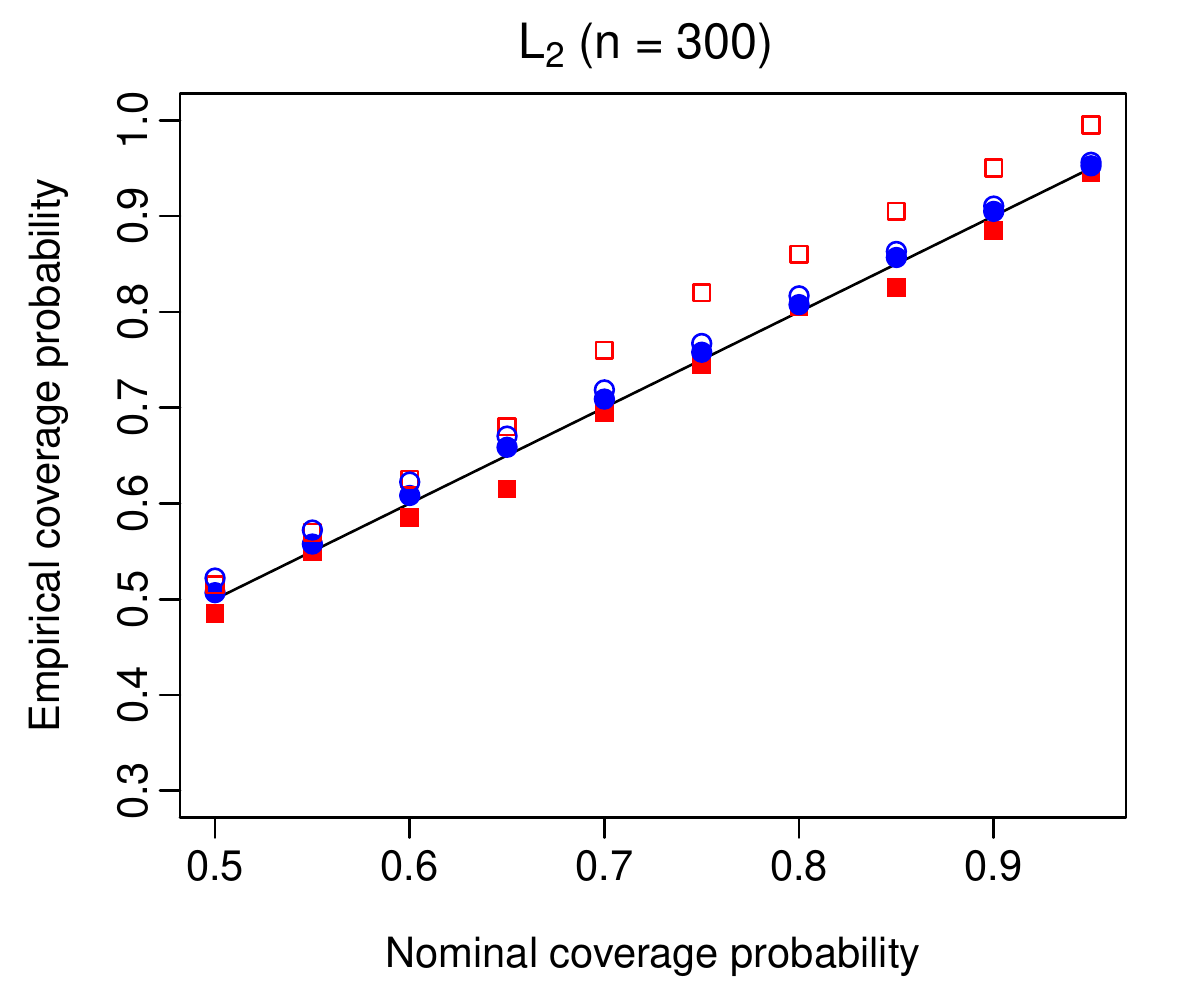}
\caption{Empirical and nominal coverage probabilities for estimating the functional trimmed mean based on the $\alpha$-radius depth, using $B_1 = B_2 =399$ repetitions and $R=200$ replications.}\label{fig:5}
\end{figure}

In Figures~\ref{fig:4} and~\ref{fig:5}, we show the nominal and empirical coverage probabilities for estimating the functional trimmed mean based on the \citeauthor{FM01}'s \citeyearpar{FM01} depth and the $\alpha$-radius depth. Similar to the functional mean, the empirical coverage probability improves for all the bootstrap procedures as the sample size increases from $n=100$ to 300. Subject to the same pseudo-random seed, the single bootstrap procedure is outperformed by the double bootstrap procedure for most, if not all, confidence levels. This result again demonstrates the advantage of the double bootstrap procedure, which can correct coverage error. Between bootstrapping the original and smoothed functions, there is an advantage to bootstrap the original function.

\begin{figure}[!htbp]
\centering
\includegraphics[width=8.5cm]{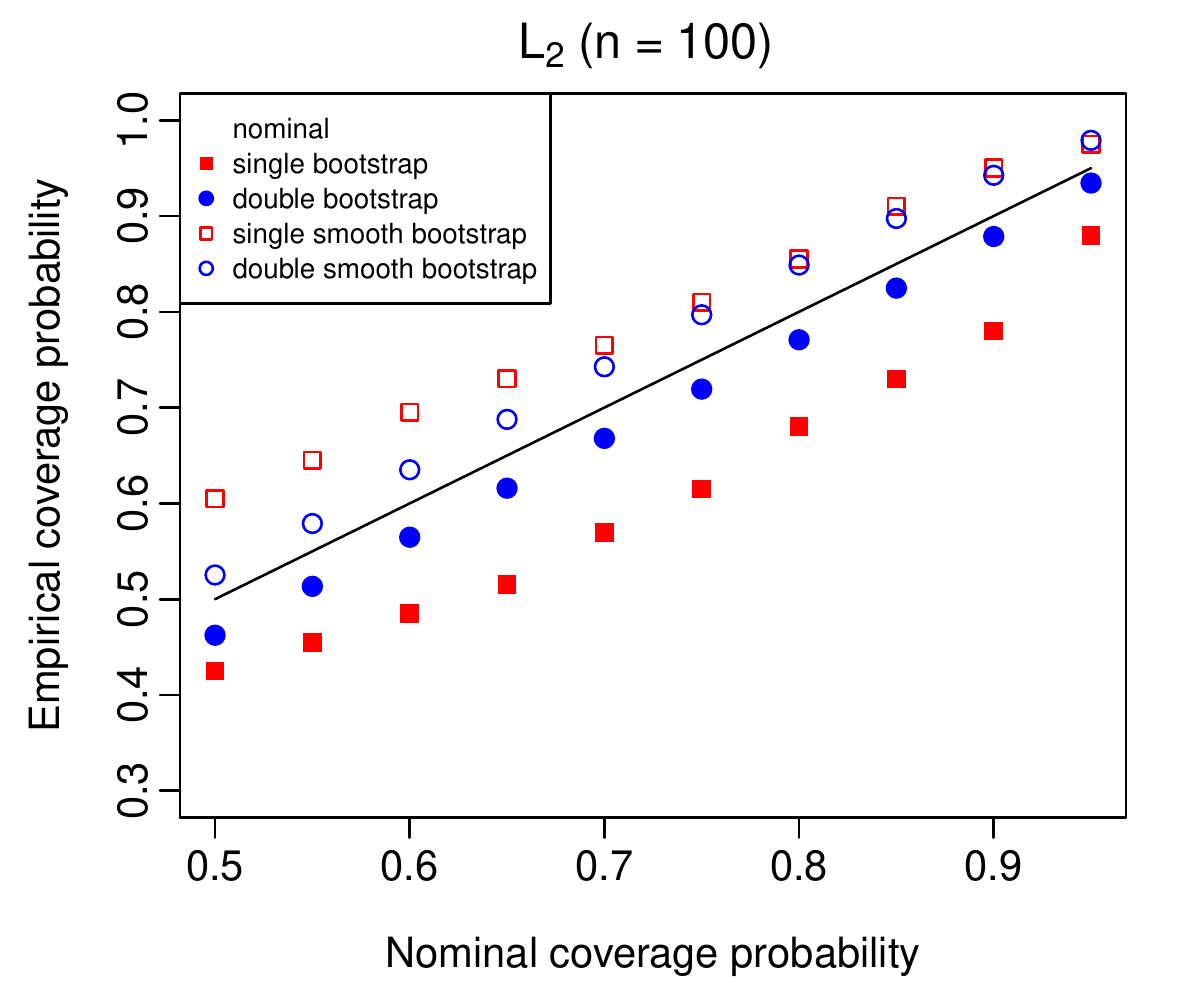}\qquad
\includegraphics[width=8.5cm]{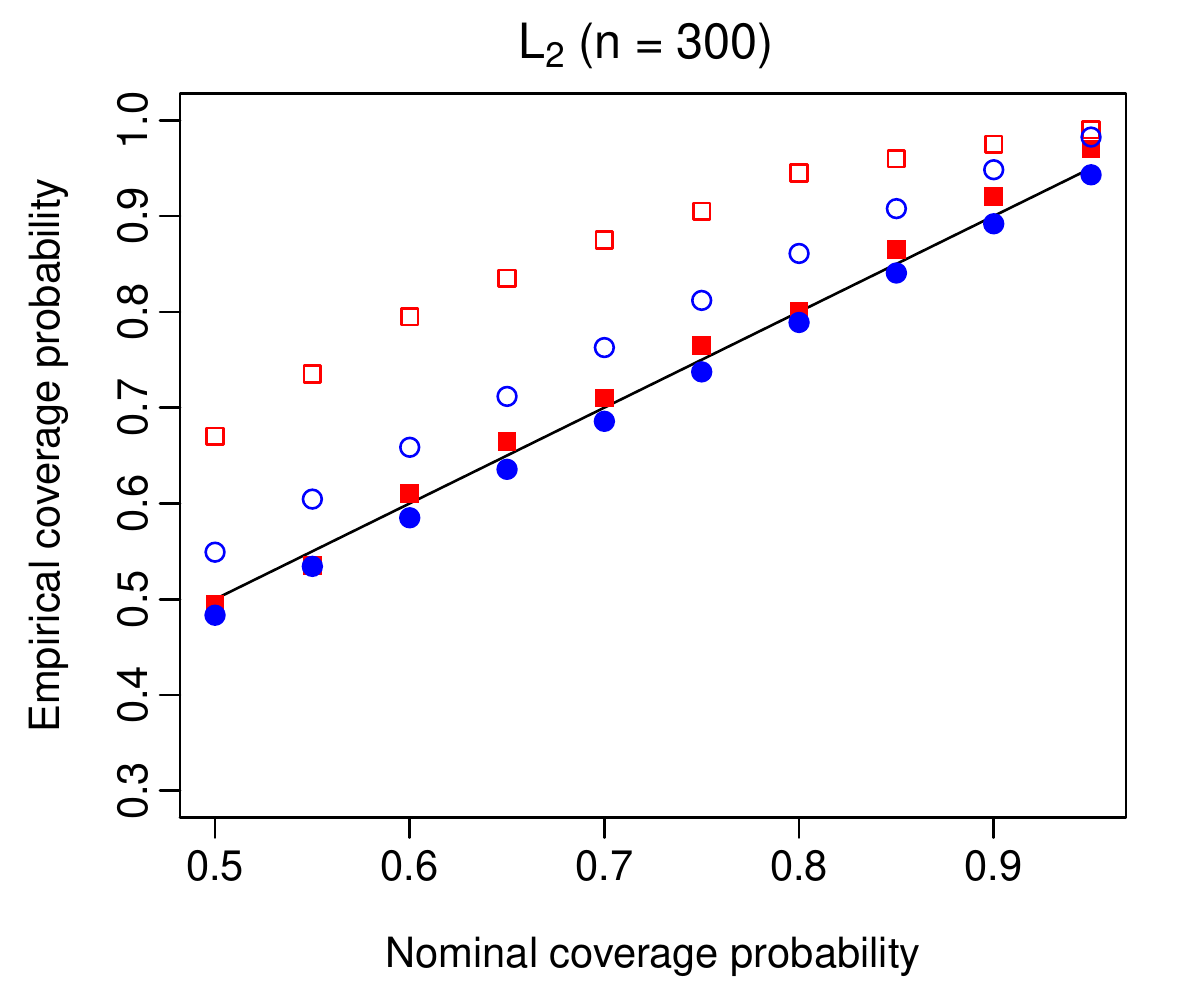}
\caption{Empirical and nominal coverage probabilities for estimating the functional variance, based on $B_1 = B_2 =399$ repetitions and $R=200$ replications.}\label{fig:6}
\end{figure}

In Figure~\ref{fig:6}, we plot the nominal and empirical coverage probabilities for estimating functional variance. Similar to the functional mean and trimmed functional mean, the empirical coverage probability improves for all the bootstrap procedures as the sample size increases from $n=100$ to 300. When $n=100$, the inferior estimation accuracy of the single bootstrap procedure is magnified, because the variance is a squared distance between $\mathcal{X}_i(t)$ and $\overline{\mathcal{X}}(t)$. When $n=300$, the estimation accuracy of the single bootstrap procedure improves, but the double bootstrap procedure still outperforms it. This result again demonstrates the usefulness of the double bootstrap procedure because of its better calibration. Between bootstrapping the original and smoothed functions, there is an advantage to bootstrap original functions.

When the descriptive statistic includes the functional mean, it is advantageous to bootstrap the original functions; when the descriptive statistic includes the functional median, it is advantageous to bootstrap the smoothed functions. In Appendix B, we perform a sensitivity analysis of bootstrap replications and examine whether or not the bootstrap replications may affect the finite sample estimation of the empirical coverage probabilities at various confidence levels. 

\section{Application to meteorology data}\label{sec:4}

We consider the classic Canadian weather station data, which is available publicly at the \textit{fda} package \citep{RWG+14}. This data set has been widely studied by \cite{RS02, RS05}, in the area of descriptive analysis and regression analysis of functional data.

\begin{figure}[!htbp]
\centering
\includegraphics[width=13cm]{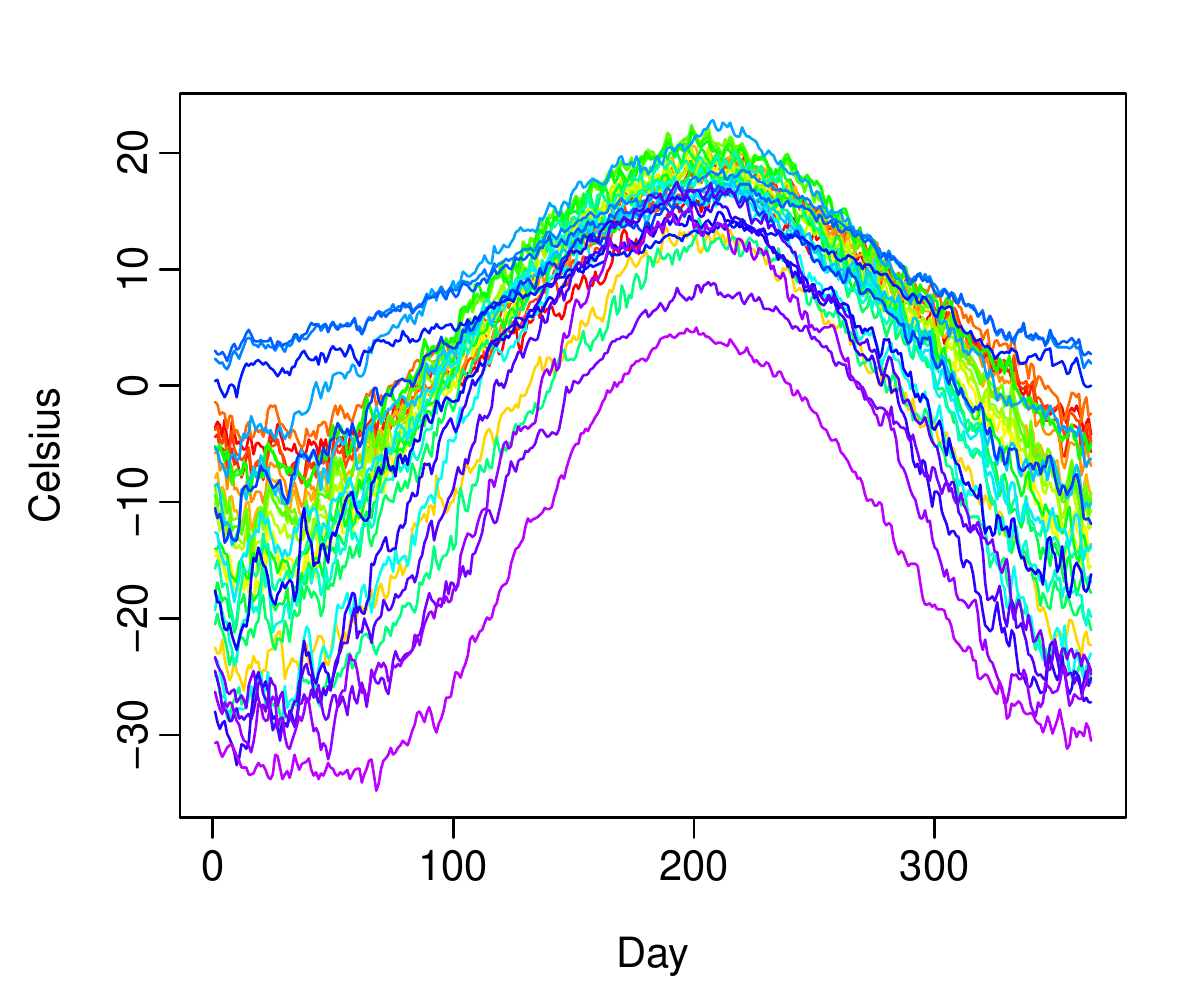}
\caption{Averaged Canadian daily temperatures from 1960 to 1994 observed at 35 weather stations. Each curve shows averaged Canadian daily temperatures at a weather station, not at a particular year.}\label{fig:7}
\end{figure}

In Figure~\ref{fig:7}, we plot the temperature change in degrees Celsius throughout a year, taken from 35 weather stations across Canada. The 35 weather stations cover the Atlantic, Pacific, Continental and Arctic climate zones. The functional curves were interpolated from 365 data points, which measure the daily mean temperature recorded by a weather station averaged over the period from 1960 to 1994. Through the rainbow plot of \cite{HS10}, the colours correspond to the geographic climates of stations; the red lines show the weather stations located in the comparably warmer regions, whereas the purple lines show the weather stations located in the colder regions \citep{Shang15}.

\begin{figure}[!htbp]
\centering
\includegraphics[width=8.5cm]{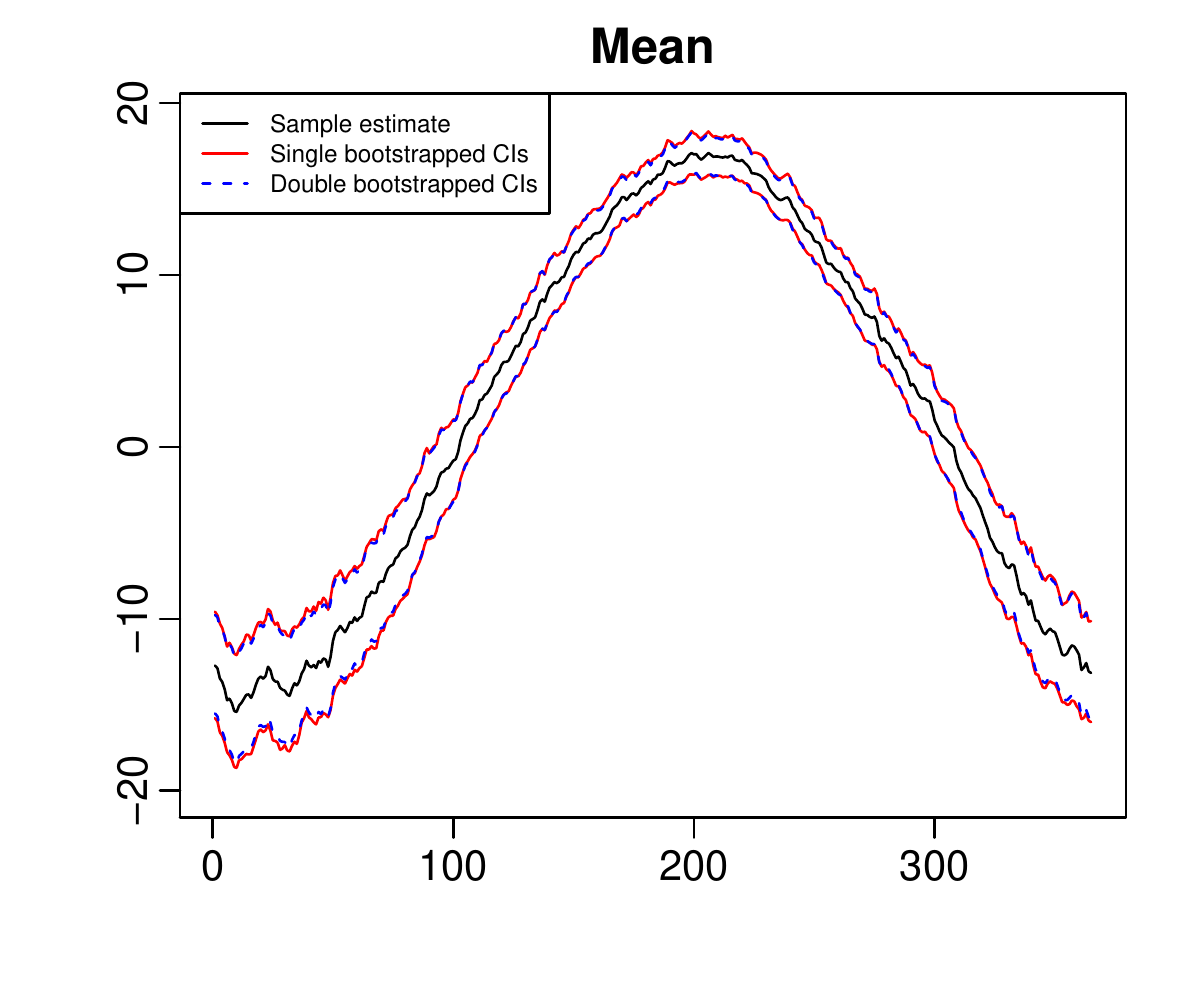}
\qquad
\includegraphics[width=8.5cm]{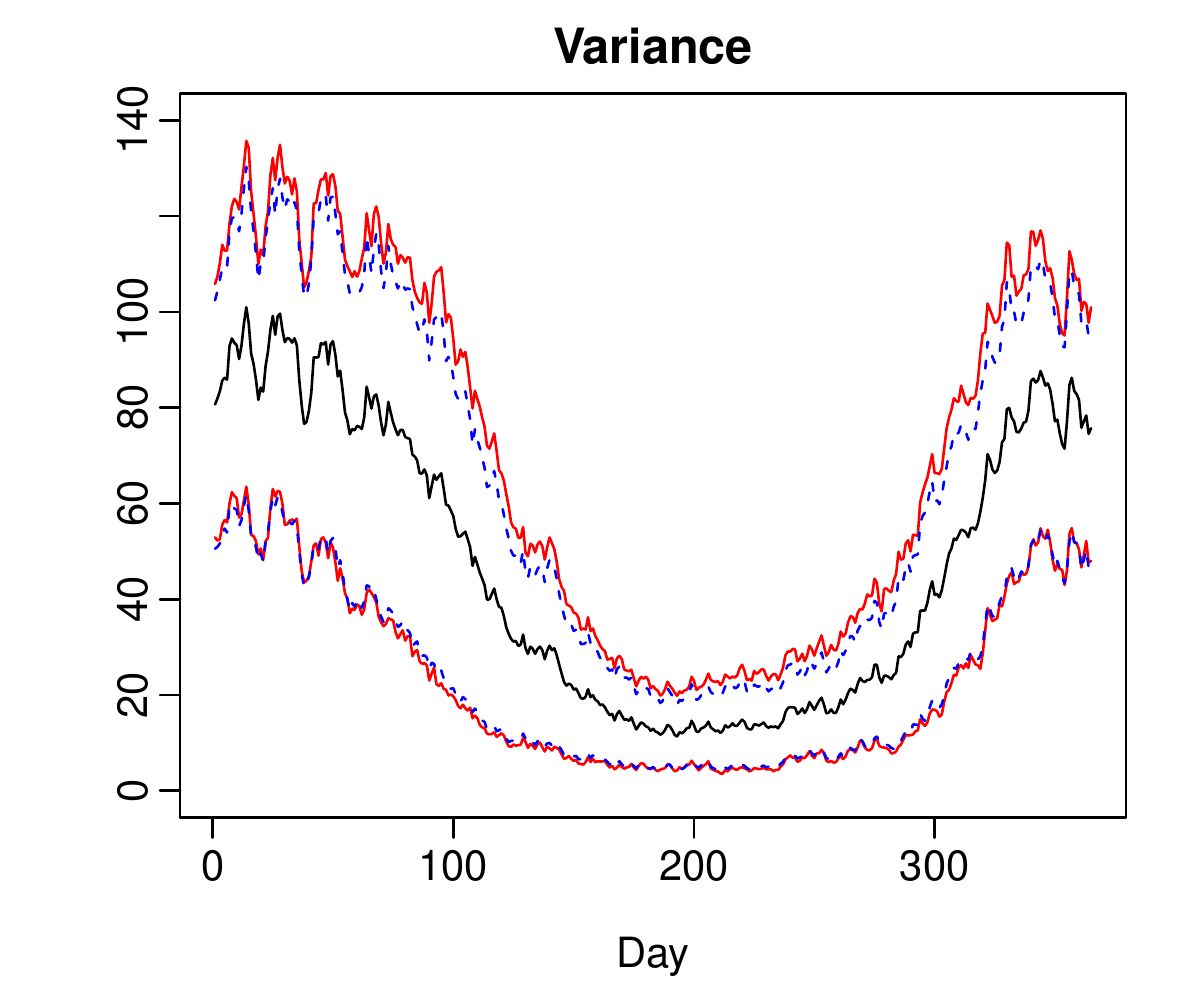}
\\
\includegraphics[width=8.5cm]{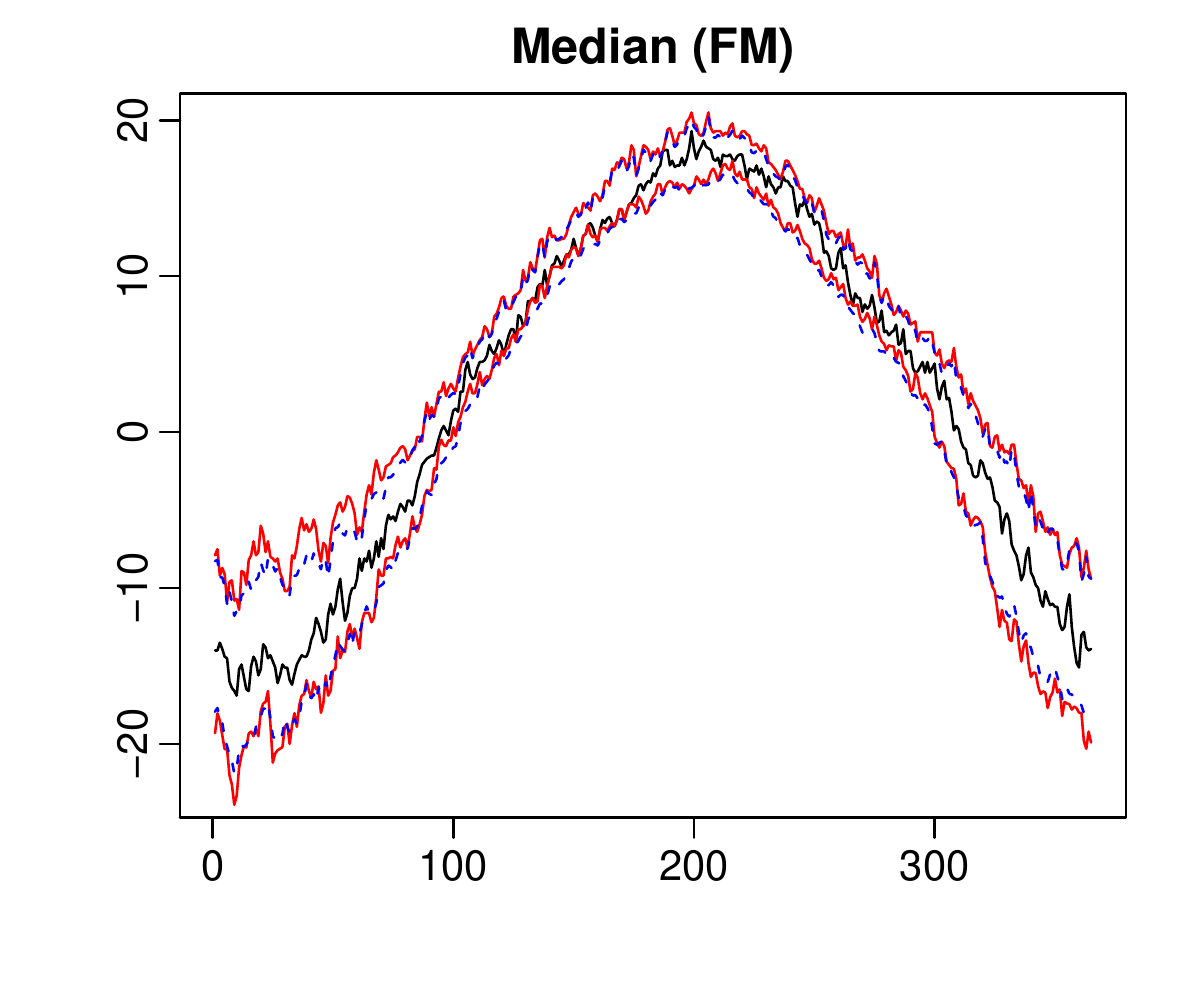}
\qquad
\includegraphics[width=8.5cm]{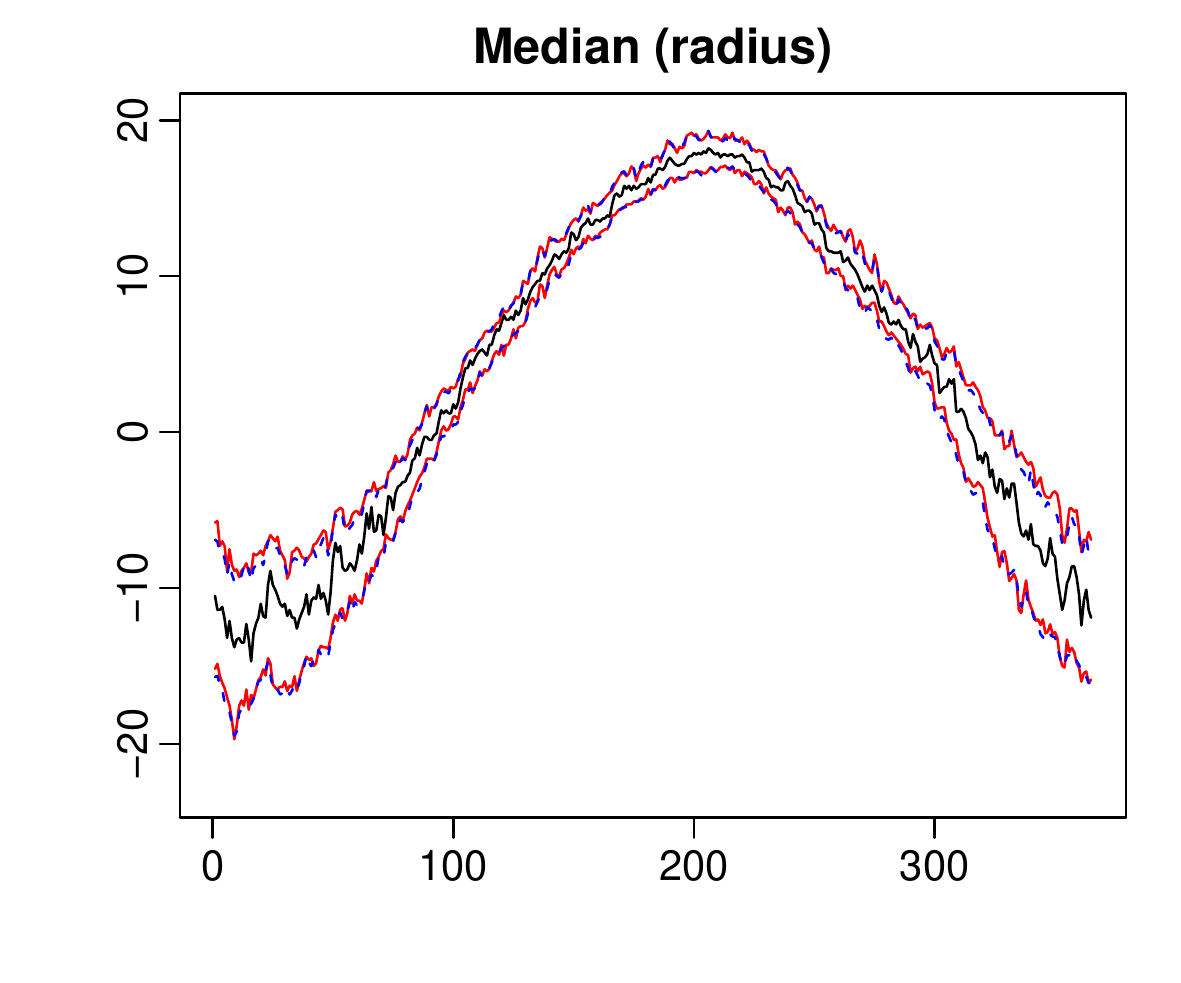}
\\
\includegraphics[width=8.5cm]{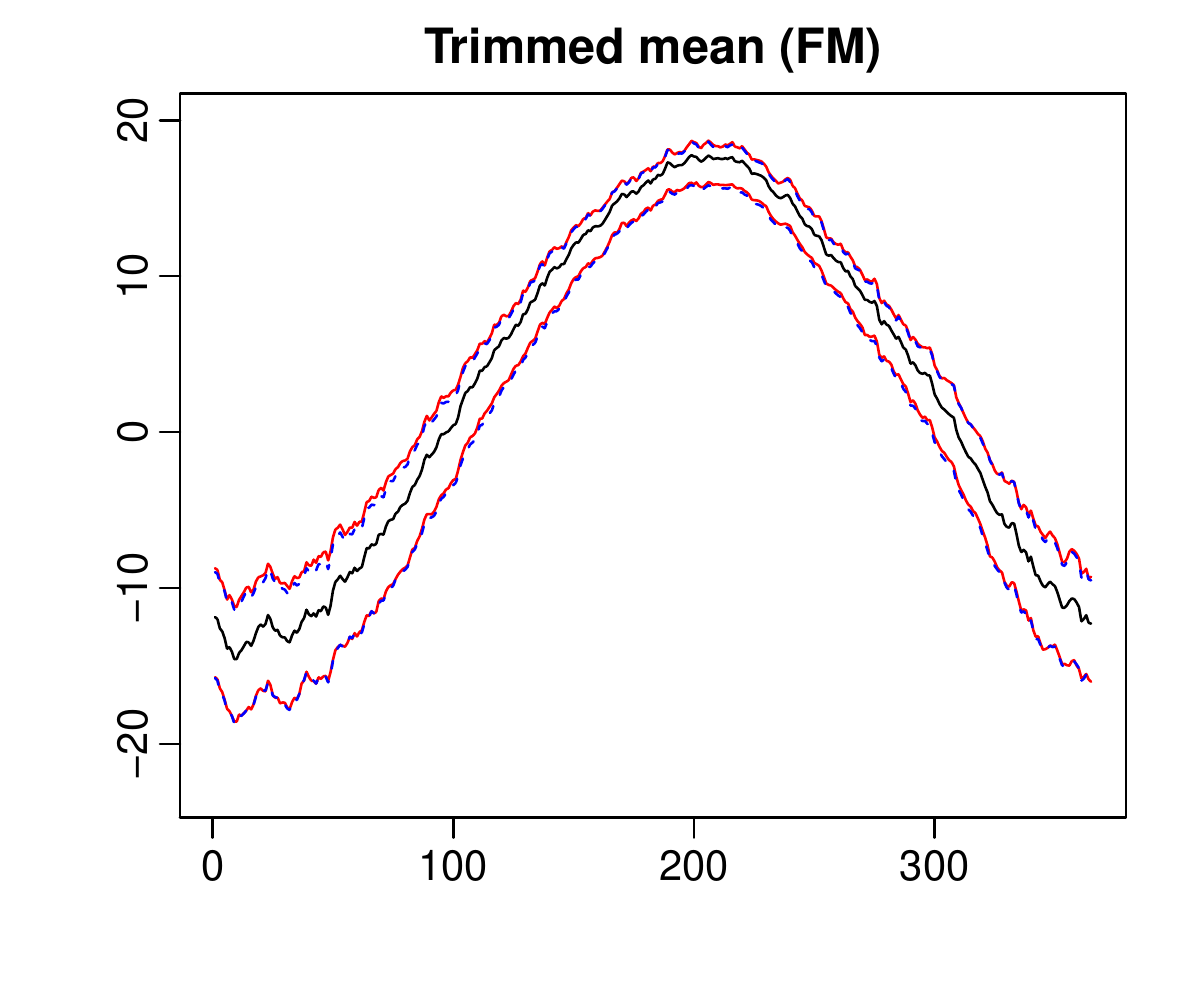}
\qquad
\includegraphics[width=8.5cm]{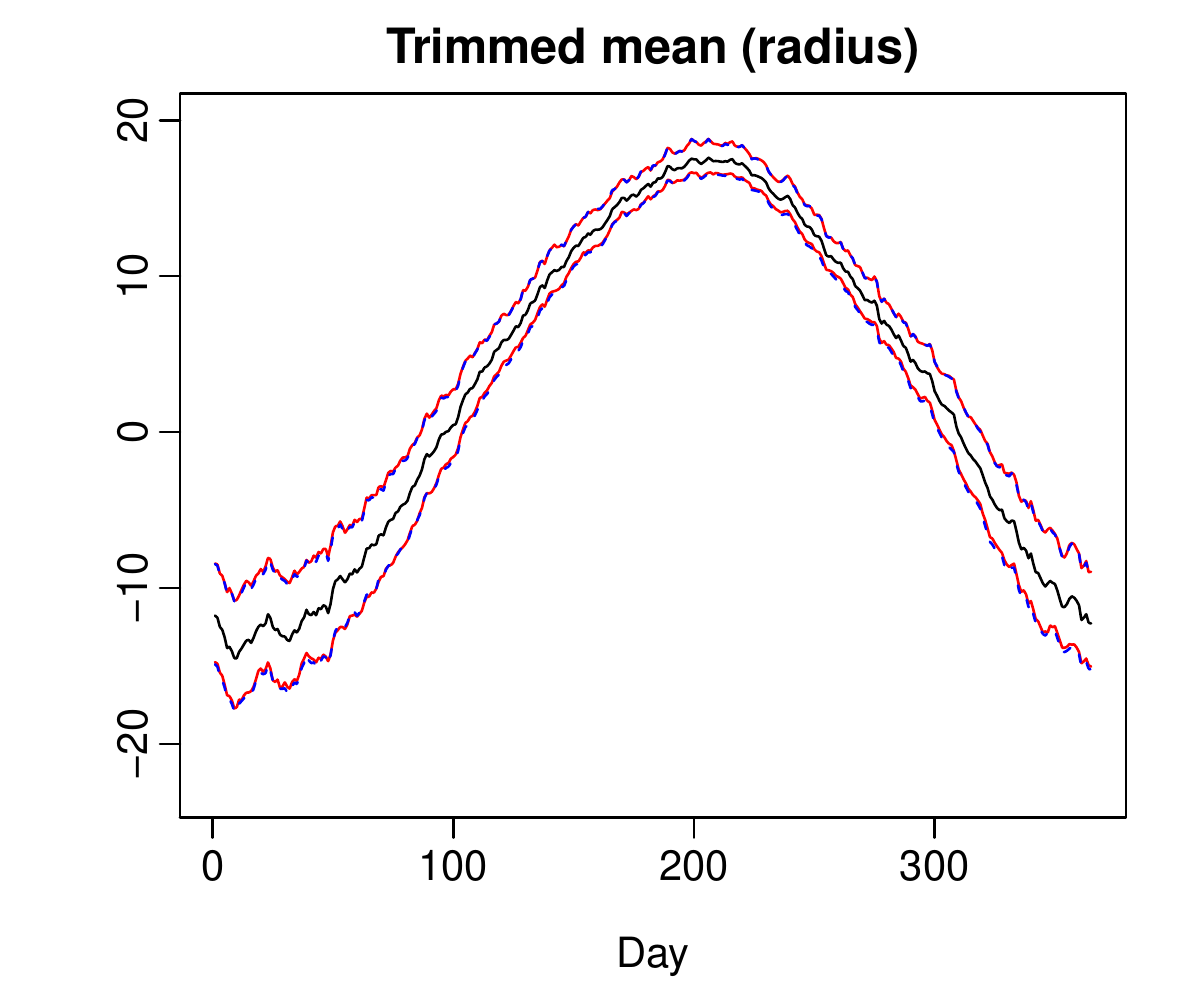}
\caption{Ninety-five percent CIs of the descriptive statistics for the Canadian weather station data, based on $B_1 = B_2 = 399$ repetitions. The sample estimates are shown in solid black lines; their 95\% CIs obtained from the single and double bootstrap procedures are shown in solid red lines and dotted blue lines, respectively.}\label{fig:8}
\end{figure}

Via the single and double bootstrap procedures, we aim to visualise the distribution of the descriptive statistics of the Canadian weather station data. As an illustration, we apply the single and double bootstrap procedures to plot the 95\% CIs of the functional mean, functional variance, functional median, and 5\% trimmed functional mean in Figure~\ref{fig:8}. The sample estimates are shown in solid black lines. While the 95\% empirical CIs are shown in solid red lines for the single bootstrap, the 95\% empirical CIs are shown in dotted blue lines for the double bootstrap.

\section{Conclusion}\label{sec:6}

We present a double bootstrap procedure for drawing random samples from a set of i.i.d. functional data to visualise the distribution of descriptive statistics. Through a series of Monte Carlo simulations, we show reduced confidence level error and improved coverage probability in comparison to the single bootstrap procedure, using the same bootstrap method. This result is not surprising since as has been pointed out by \cite{CH15} and \cite{Shang15}, iterating the bootstrap principle reduces the dependence between the probability distribution of the resample and the unknown data generating process. As the number of sample curves increases from 100 to 300, the estimation accuracy improves for both bootstrap procedures, more so for the double bootstrap. Illustrated by the Canadian weather station data set, the single and double bootstrap procedures produce similar results for estimating the distributions of various descriptive statistics, but there are noticeable differences at the edges of function support for the functional median.

There are a few ways in which the paper can be further extended, and we briefly outline two. Firstly, bootstrapping functional time series is still in its infancy, and we intend to extend the double bootstrap procedure to analyse stationary and weakly dependent functional time series \citep[see, e.g.,][for various single bootstrap procedures]{ZP16, FN16, Shang16, Paparoditis16, PS20}. Secondly, we aim to develop bootstrap procedures that can handle stationary long-memory functional time series \citep[see, e.g.,][]{LRS17} and non-stationary long-memory functional time series \citep[see, e.g.,][]{LRS20}.


\vspace{-.15in}

\section*{Acknowledgement}

The author thanks Dr. Yanrong Yang for many comments and suggestions.

\vspace{-.15in}


\begin{appendix}
\section*{Appendix A: $L_{\infty}$ distance metrics}
\vspace{.1in}

Similar to Section~\ref{sec:33} of the main manuscript, we also consider $L_{\infty}$ distance metrics for constructing CIs. The $L_{\infty}$ distance metrics are defined as
\begin{align*}
\left\|\theta(t) - \widehat{\theta}(t)\right\|_{\infty} &= \sup_{t\in\mathcal{I}} \left|\theta(t) - \widehat{\theta}(t)\right|, \\
\left\|\widehat{\theta}(t) - \widehat{\theta}^b(t)\right\|_{\infty} &= \sup_{t\in\mathcal{I}} \left|\widehat{\theta}(t) - \widehat{\theta}^b(t)\right|, \\
\left\|\widehat{\theta}^{b}(t) - \widehat{\theta}^{b\eta}(t)\right\|_{\infty} &= \sup_{t\in\mathcal{I}} \left|\widehat{\theta}^{b}(t) - \widehat{\theta}^{b\eta}(t)\right|.
\end{align*}

Using the $L_{\infty}$ distance metrics for constructing CIs, Figures~\ref{fig:Appendix_A_1} to~\ref{fig:Appendix_A_6} present the finite sample performance between the two bootstrap procedures. 

\renewcommand\thefigure{A.\arabic{figure}}    
\setcounter{figure}{0}
\begin{figure}[!htbp]
\centering
\includegraphics[width=8.5cm]{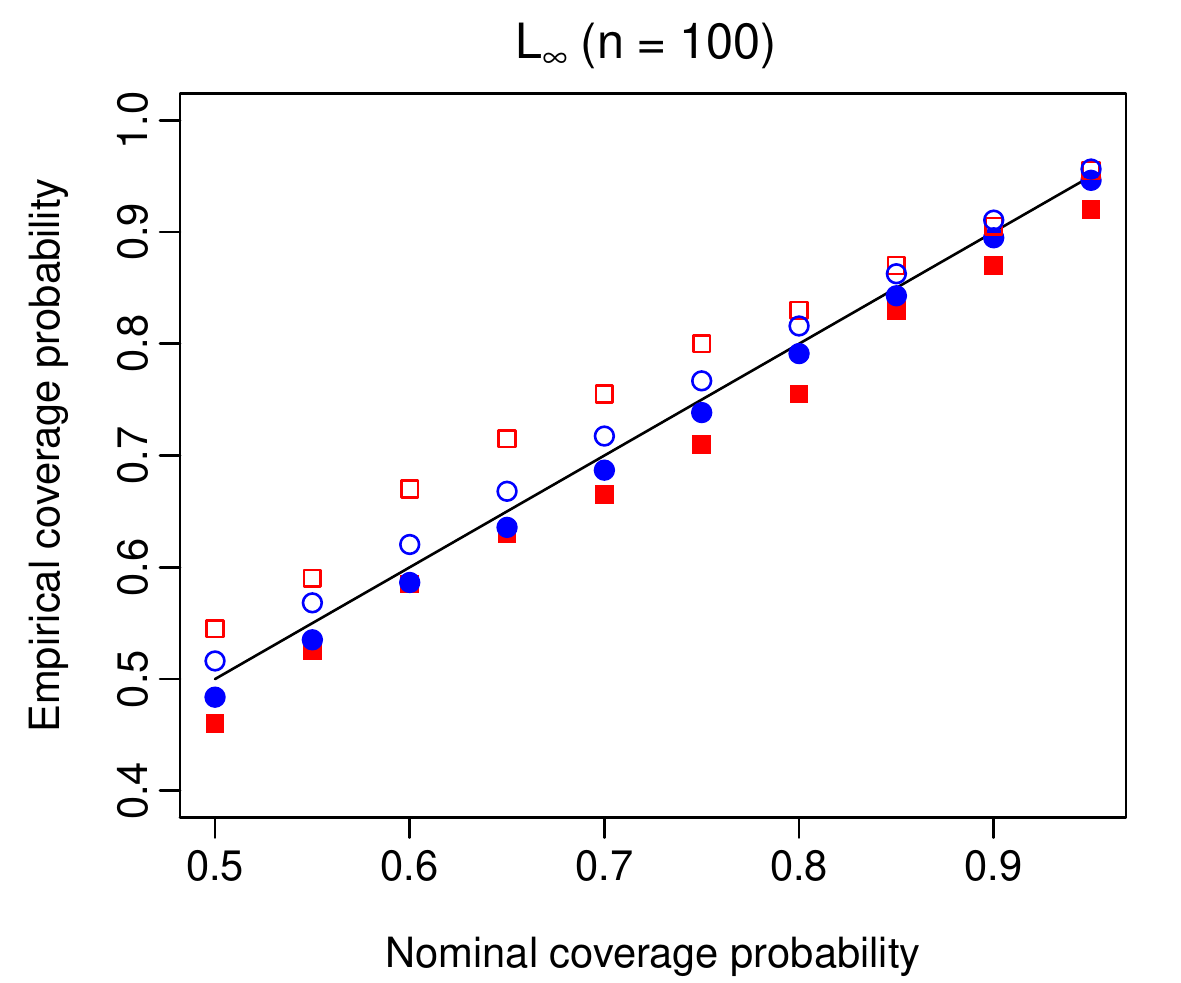}
\qquad
\includegraphics[width=8.5cm]{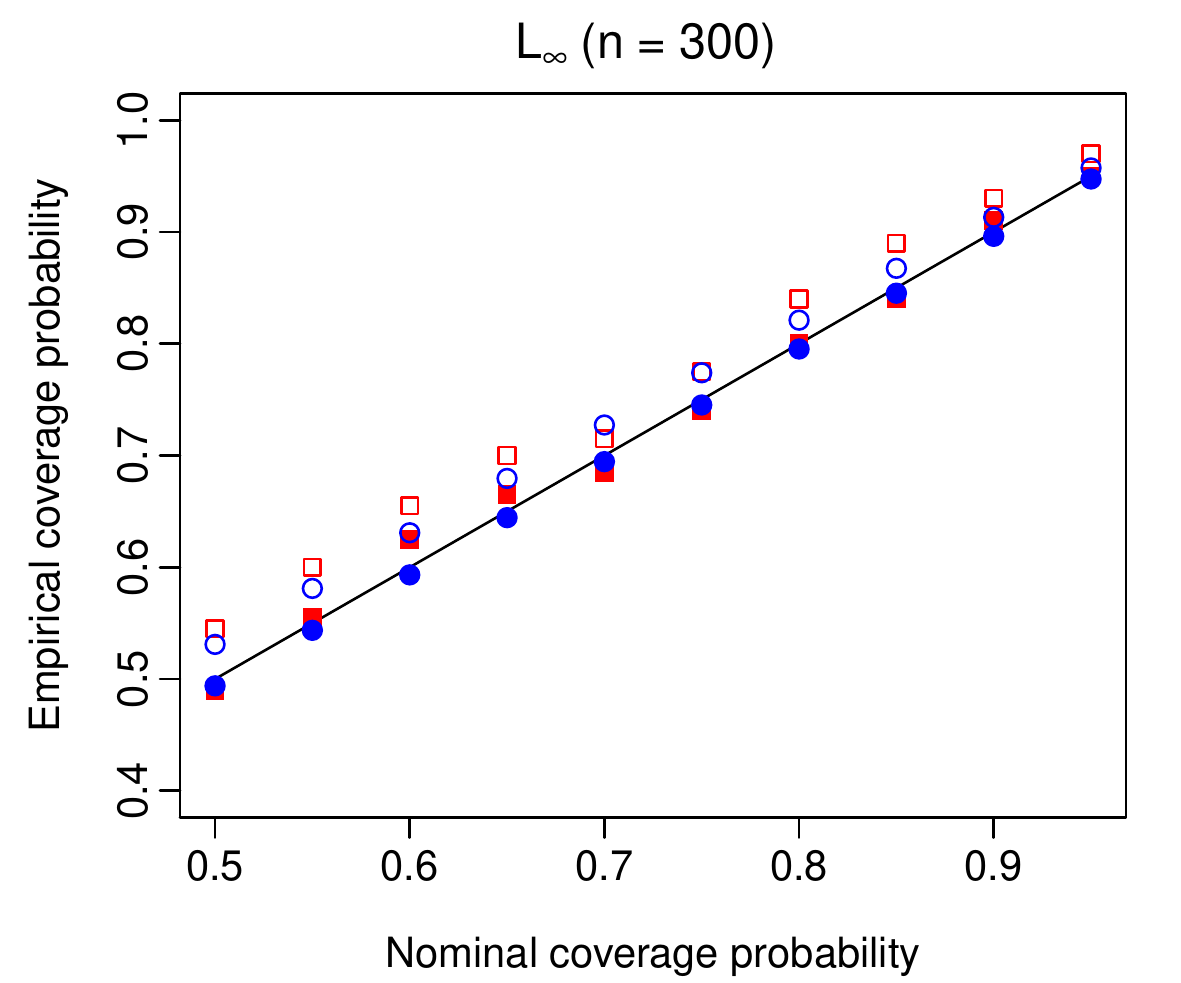}
\caption{Empirical and nominal coverage probabilities for estimating the functional mean, based on $B_1 = B_2 =399$ repetitions and $R=200$ replications. See~\eqref{eq:smoothness} for the smooth bootstrap.}\label{fig:Appendix_A_1}
\end{figure}

\begin{figure}[!htbp]
\centering
\includegraphics[width=8.5cm]{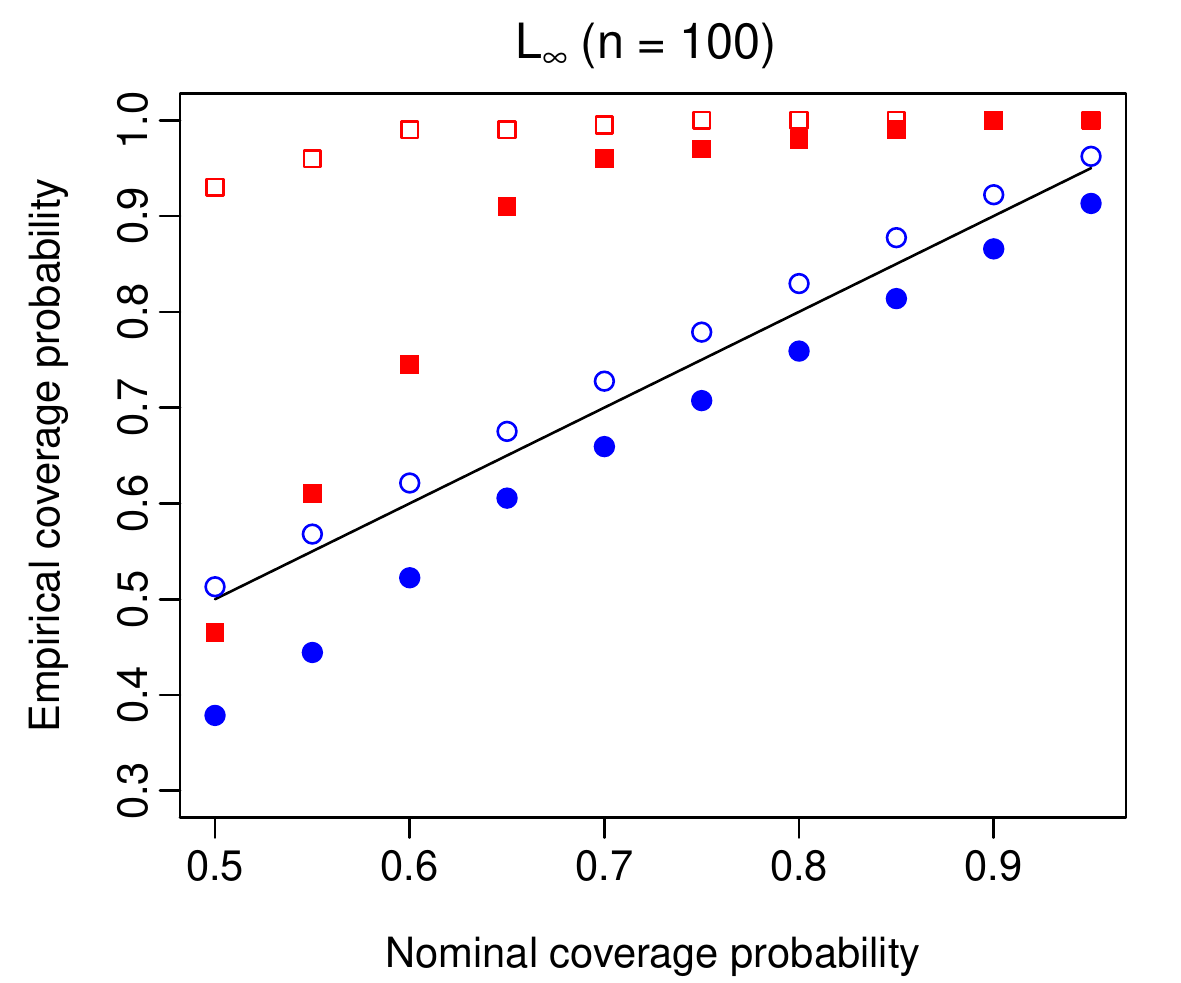}
\qquad
\includegraphics[width=8.5cm]{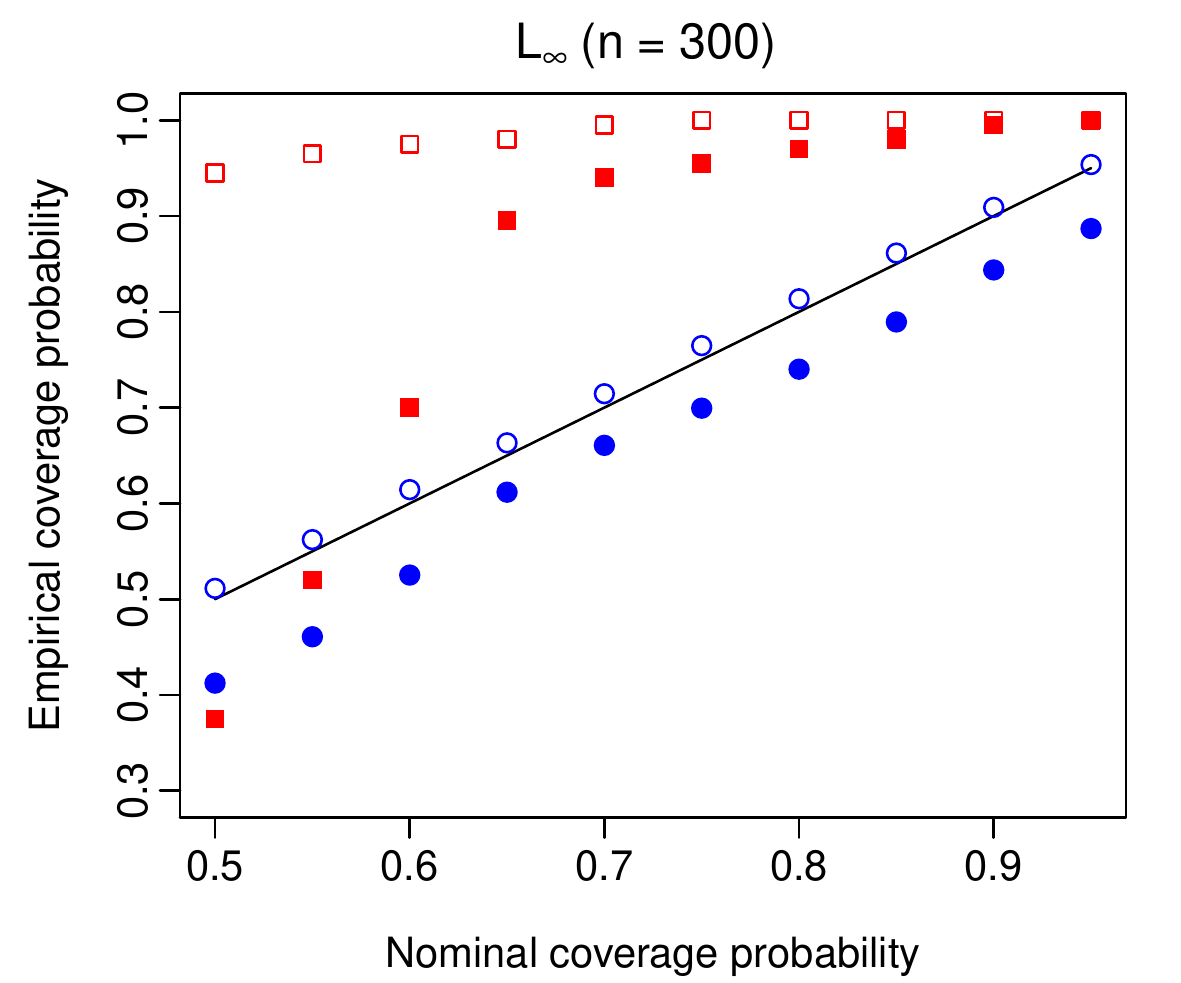}
\caption{Empirical and nominal coverage probabilities for estimating the functional median based on the \citeauthor{FM01}'s \citeyearpar{FM01} depth, using $B_1 = B_2 =399$ repetitions and $R=200$ replications.}\label{fig:Appendix_A_2}
\end{figure}

\begin{figure}[!htbp]
\centering
\includegraphics[width=8.5cm]{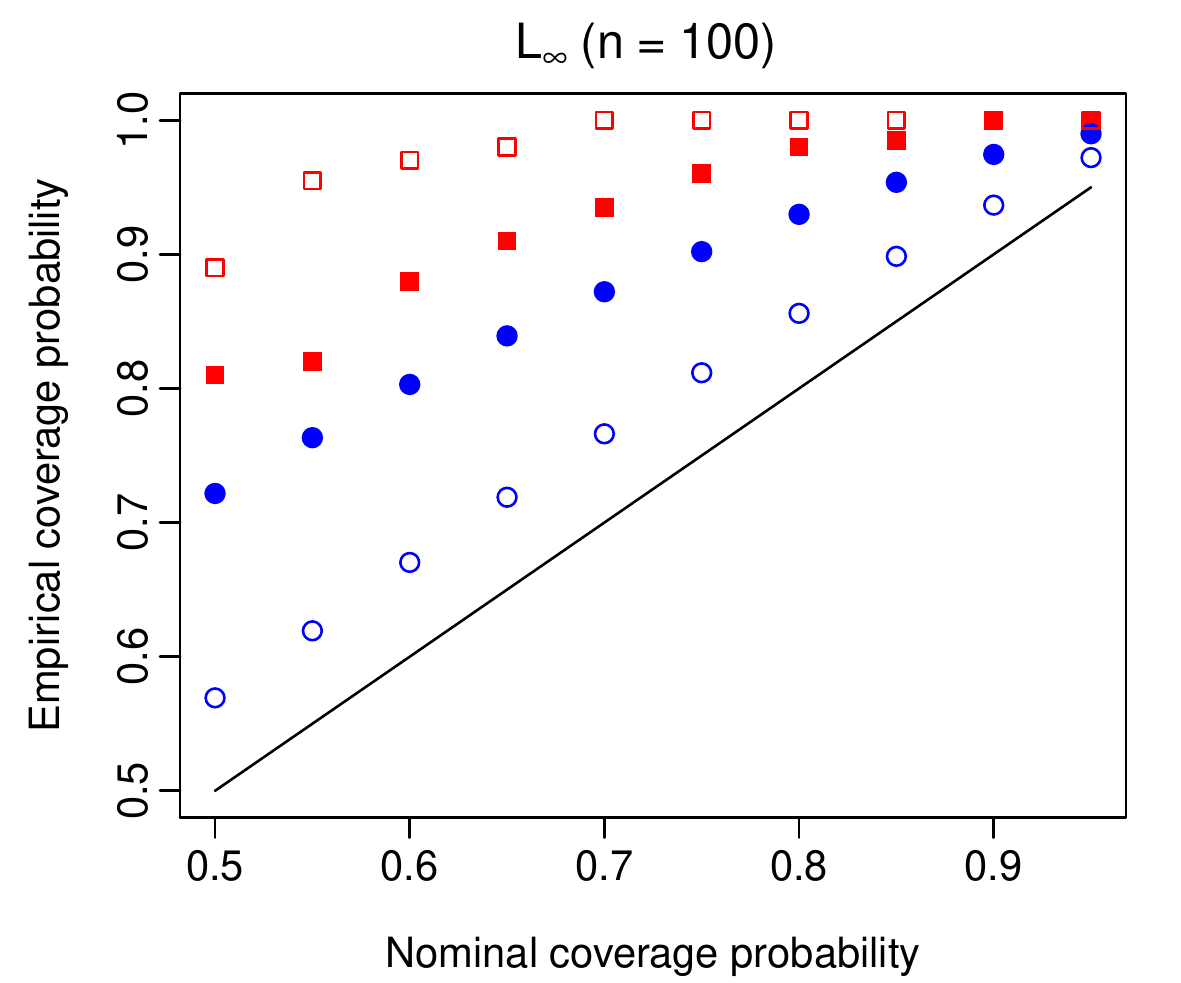}
\qquad
\includegraphics[width=8.5cm]{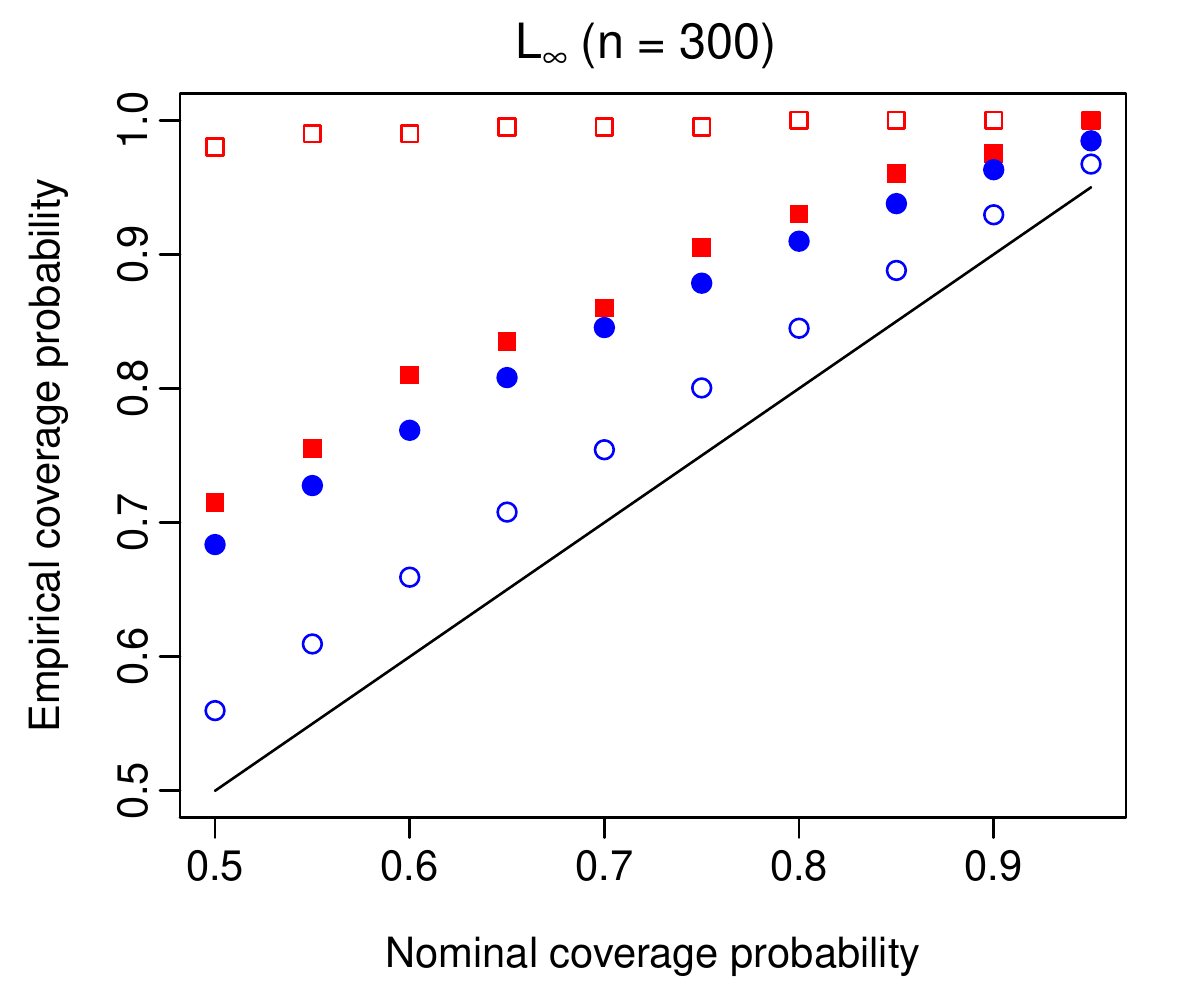}
\caption{Empirical and nominal coverage probabilities for estimating the functional median based on the $\alpha$-radius depth, using $B_1 = B_2 =399$ repetitions and $R=200$ replications.}\label{fig:Appendix_A_3}
\end{figure}

\begin{figure}[!htbp]
\centering
\includegraphics[width=8.5cm]{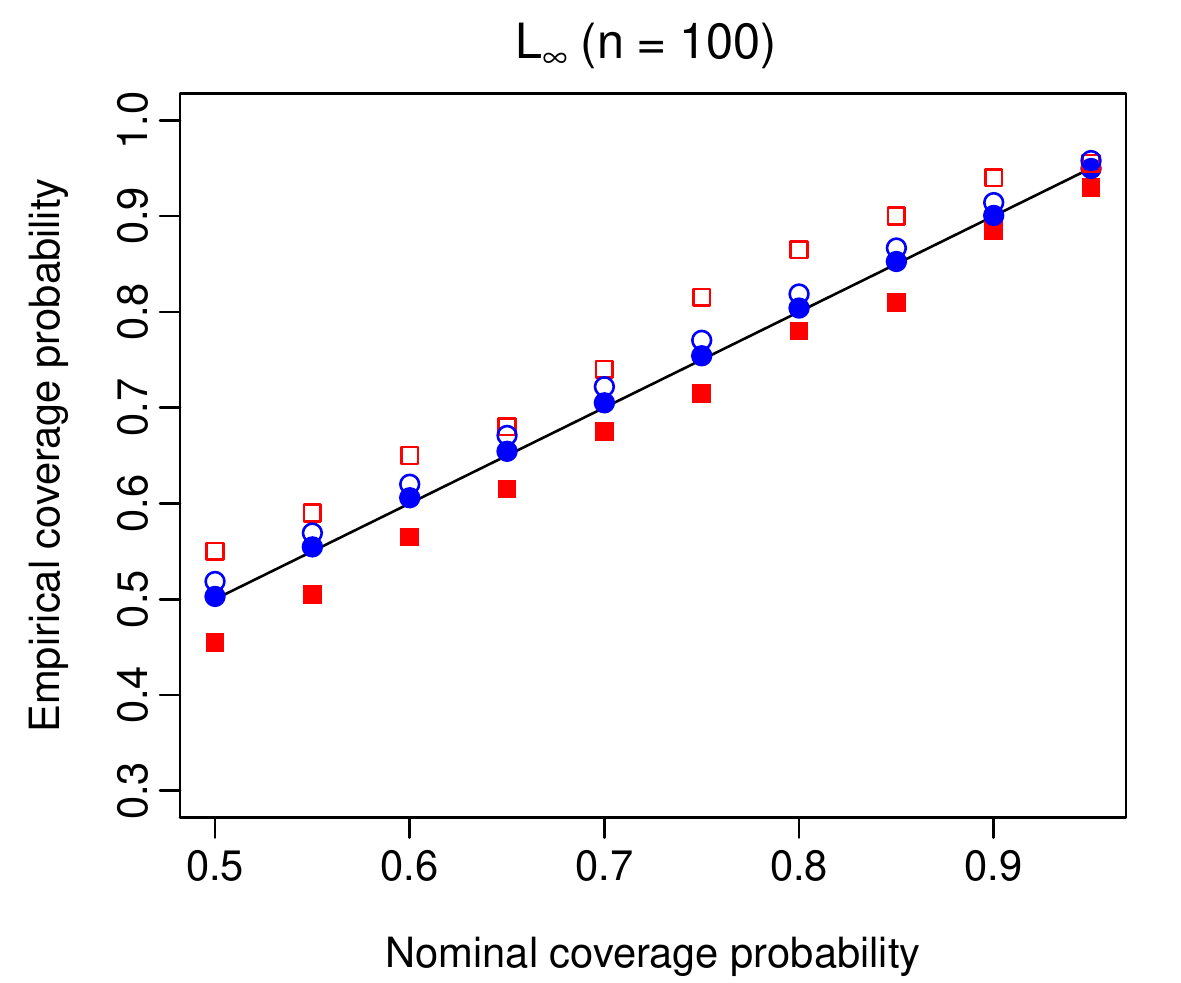}
\qquad
\includegraphics[width=8.5cm]{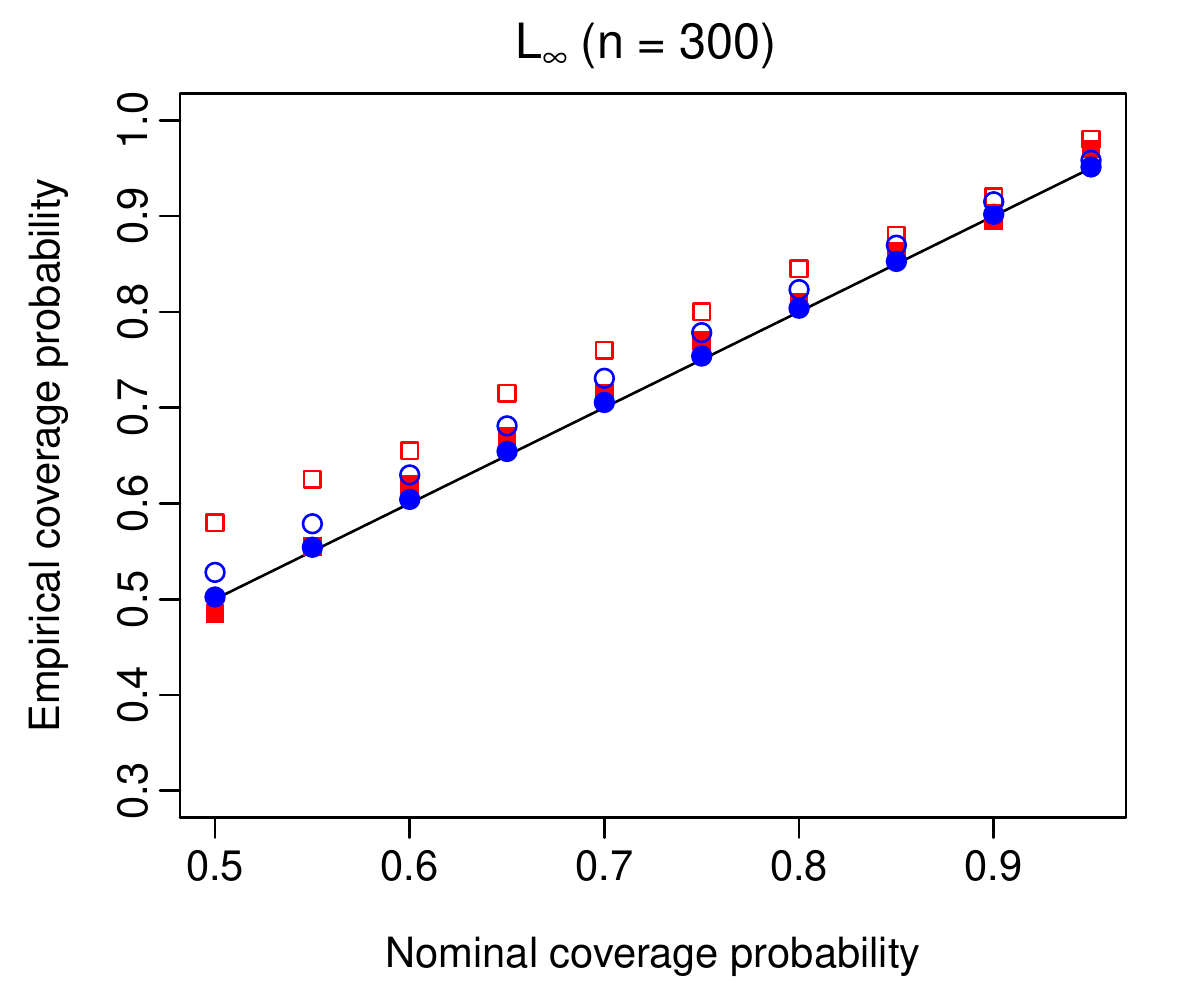}
\caption{Empirical and nominal coverage probabilities for estimating the functional trimmed mean based on the \citeauthor{FM01}'s \citeyearpar{FM01} depth, using $B_1 = B_2 =399$ repetitions and $R=200$ replications.}\label{fig:Appendix_A_4}
\end{figure}

\begin{figure}[!htbp]
\centering
\includegraphics[width=8.5cm]{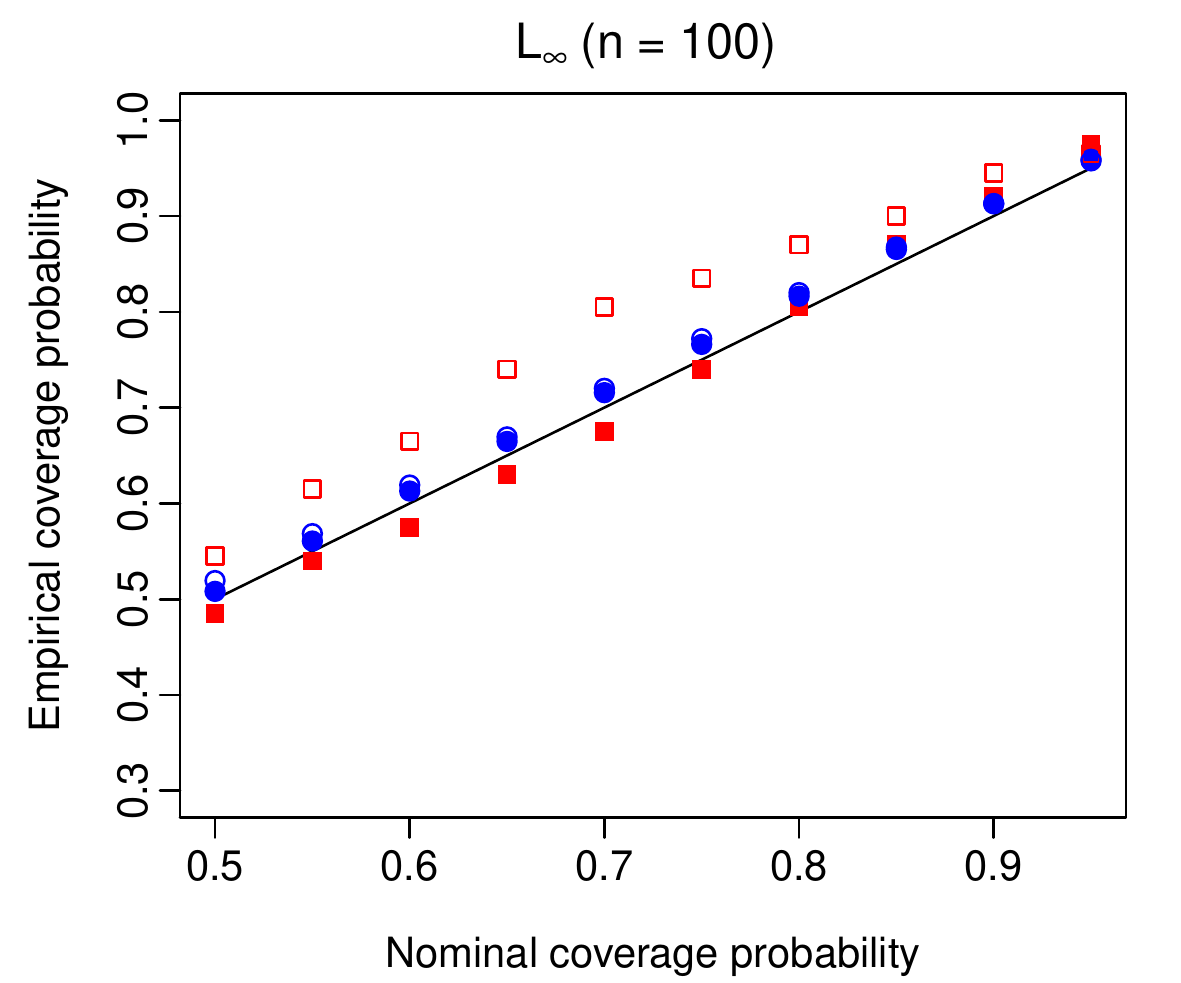}
\qquad
\includegraphics[width=8.5cm]{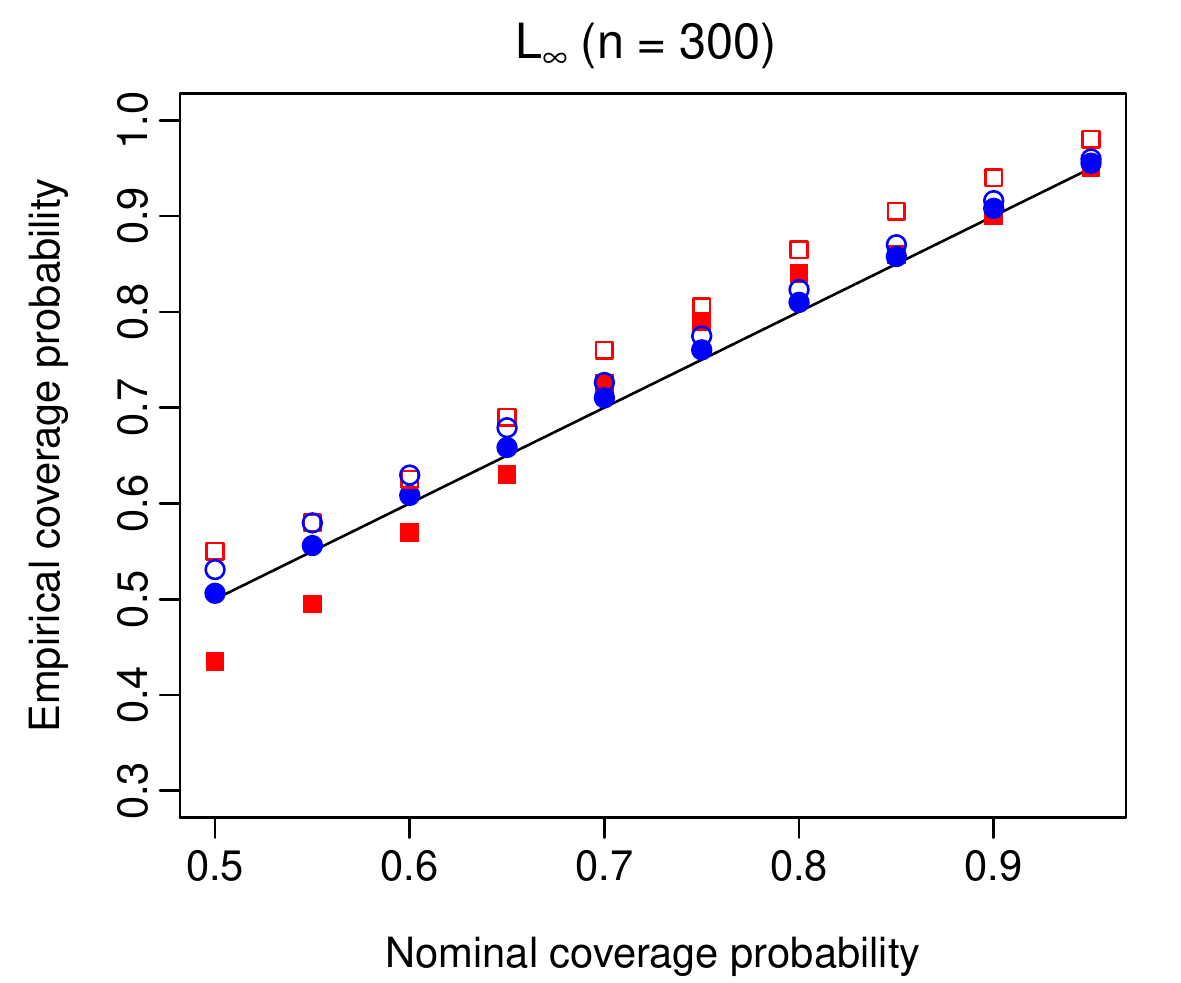}
\caption{Empirical and nominal coverage probabilities for estimating the functional trimmed mean based on the $\alpha$-radius depth, using $B_1 = B_2 =399$ repetitions and $R=200$ replications.}\label{fig:Appendix_A_5}
\end{figure}

\begin{figure}[!htbp]
\centering
\includegraphics[width=8.5cm]{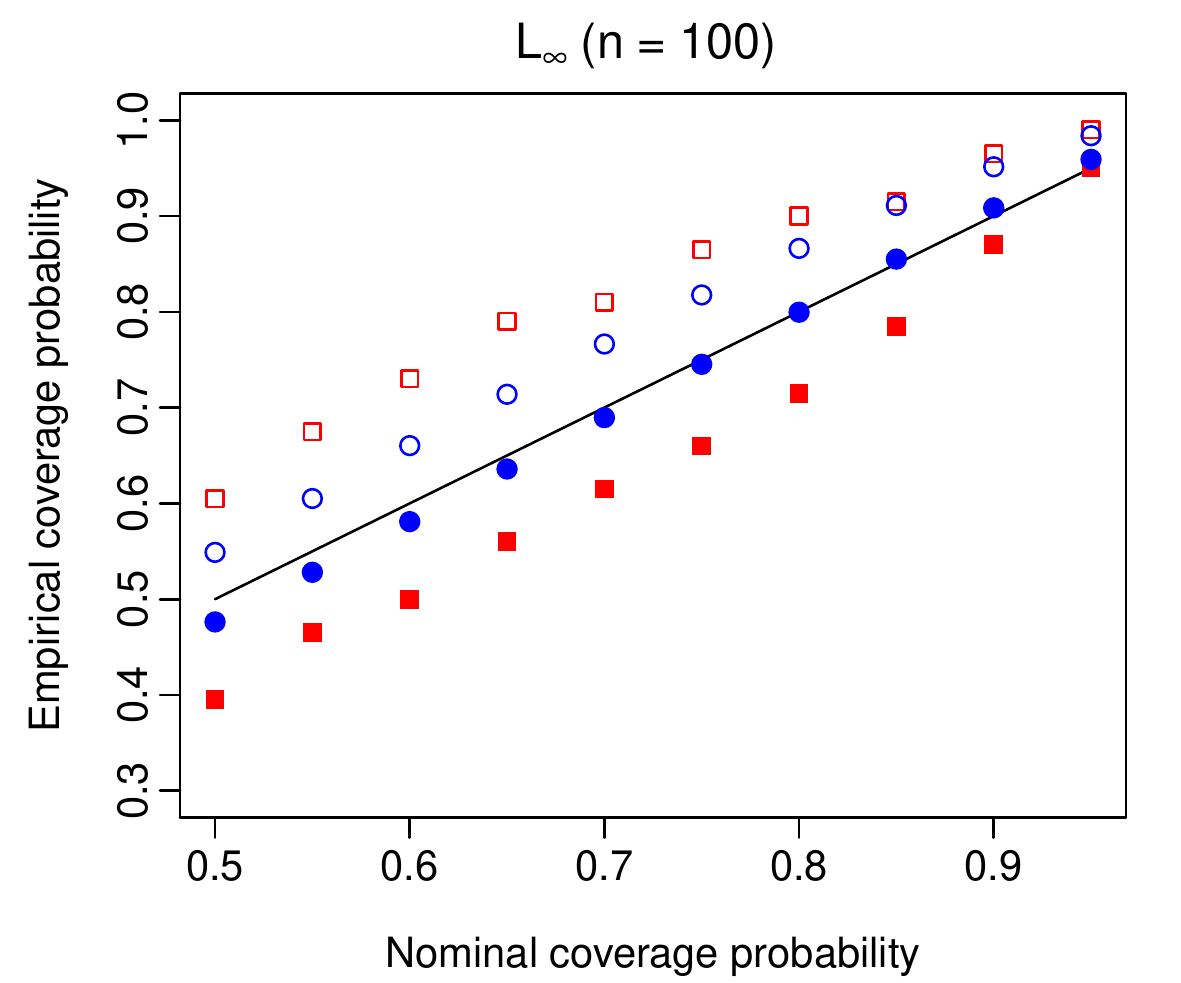}
\qquad
\includegraphics[width=8.5cm]{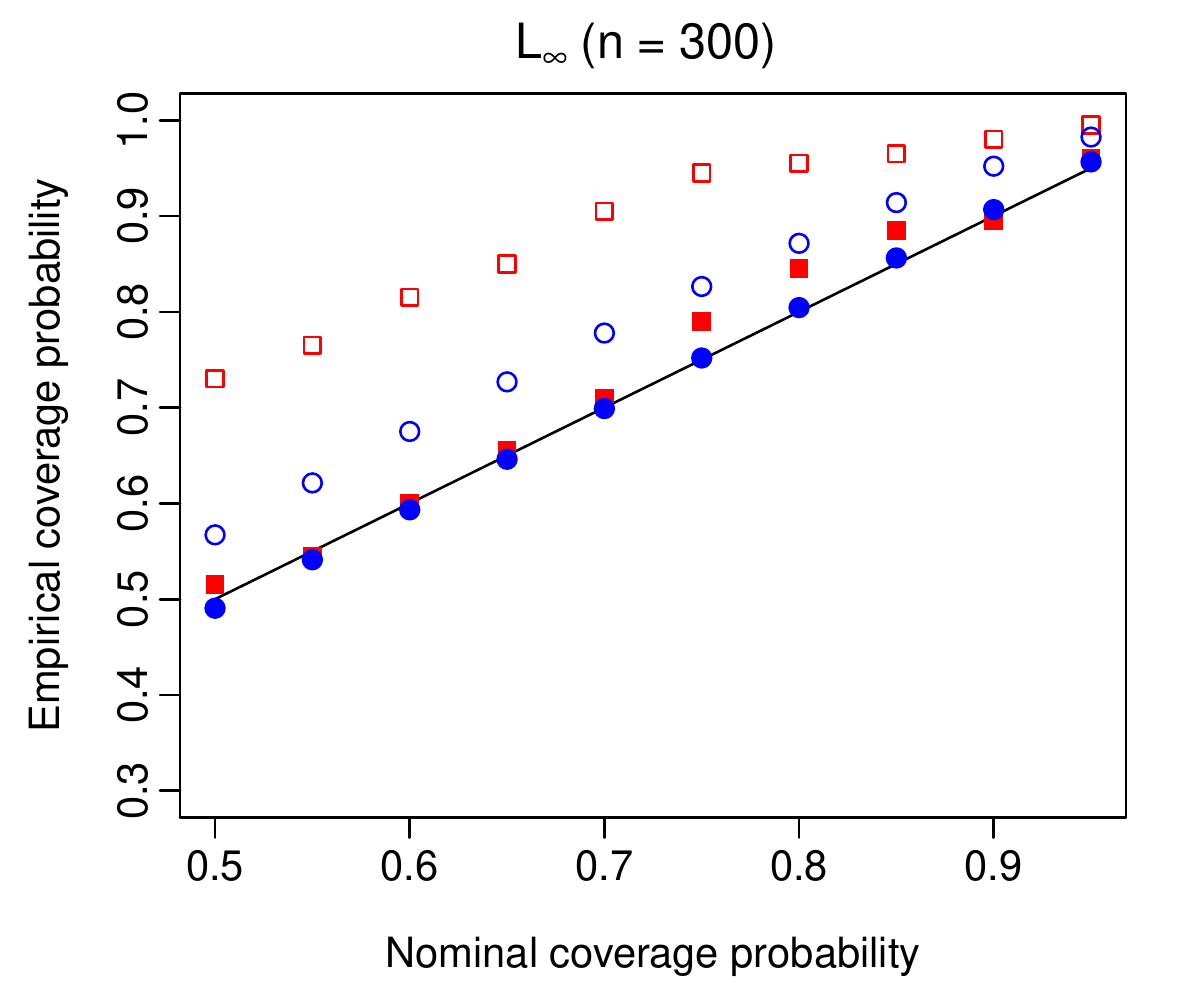}
\caption{Empirical and nominal coverage probabilities for estimating the functional variance, based on $B_1 = B_2 =399$ repetitions and $R=200$ replications.}\label{fig:Appendix_A_6}
\end{figure}


\section*{Appendix B: Sensitivity analysis of bootstrap replications}

In the double bootstrap procedure, the choices of $B_1$ and $B_2$ may affect the empirical coverage probability. In Table~\ref{tab:sensitivity}, we study four different choices of $B_1$ and $B_2$ and find that the empirical coverage probabilities are fairly similar regardless different combinations of $B_1$ and $B_2$.

\renewcommand\thetable{B.\arabic{table}}    
\setcounter{table}{0}

\vspace{-.11in}
\begin{small}
\begin{center}
\tabcolsep 0.085in
 \setlength\LTcapwidth{\hsize}
  \advance\LTcapwidth by 0pt
\begin{longtable}{@{}lllrrrrrrrrrr@{}}
\caption{Empirical and nominal coverage probabilities are shown for estimating various descriptive statistics, based on four combinations of bootstrap replications. $B_1$ symbolizes the number of bootstrap samples in the first layer of the double bootstrap, whereas $B_2$ symbolizes the number of bootstrap samples in the second layer of the double bootstrap. FM denotes the \citeauthor{FM01}'s \citeyearpar{FM01} depth.}\label{tab:sensitivity} \\
\toprule
& &  & \multicolumn{10}{c}{Nominal coverage probability} \\
Statistic & &  & 0.5 & 0.55 & 0.6 & 0.65 & 0.7 & 0.75 & 0.8 & 0.85 & 0.9 & 0.95 \\ 
\endfirsthead
\toprule
& &  & \multicolumn{10}{c}{Nominal coverage probability} \\
Statistic & &  & 0.5 & 0.55 & 0.6 & 0.65 & 0.7 & 0.75 & 0.8 & 0.85 & 0.9 & 0.95 \\ 
\midrule
\endhead
\endfoot
\endlastfoot
\midrule
Mean & $B_1=399$  &  L2 $(n=100)$ & 0.49 & 0.54 & 0.59 & 0.64 & 0.69 & 0.74 & 0.79 & 0.84 & 0.89 & 0.95 \\ 
& $B_2 = 399$ &  Linf $(n=100)$ & 0.49 & 0.54 & 0.59 & 0.64 & 0.69 & 0.74 & 0.79 & 0.84 & 0.90 & 0.95 \\ 
& &   L2 $(n=300)$ & 0.50 & 0.55 & 0.60 & 0.65 & 0.70 & 0.75 & 0.80 & 0.84 & 0.90 & 0.95 \\ 
& &   Linf $(n=300)$ & 0.49 & 0.54 & 0.59 & 0.64 & 0.69 & 0.74 & 0.80 & 0.85 & 0.90 & 0.95 \\ \\
& $B_1=399$  &  L2 $(n=100)$ & 0.48 & 0.53 & 0.58 & 0.63 & 0.68 & 0.73 & 0.78 & 0.84 & 0.89 & 0.94 \\ 
& $B_2 = 99$ &   Linf $(n=100)$ & 0.48 & 0.53 & 0.58 & 0.63 & 0.68 & 0.73 & 0.79 & 0.84 & 0.89 & 0.94 \\ 
& &   L2 $(n=300)$  & 0.49 & 0.54 & 0.59 & 0.64 & 0.69 & 0.74 & 0.79 & 0.84 & 0.89 & 0.94 \\ 
& &   Linf $(n=300)$ & 0.49 & 0.54 & 0.59 & 0.64 & 0.69 & 0.74 & 0.79 & 0.84 & 0.89 & 0.94 \\ \\
& $B_1=99$  &  L2 $(n=100)$ & 0.49 & 0.54 & 0.59 & 0.64 & 0.69 & 0.74 & 0.79 & 0.84 & 0.89 & 0.95 \\ 
& $B_2 = 399$&   Linf $(n=100)$ & 0.49 & 0.54 & 0.59 & 0.64 & 0.69 & 0.74 & 0.79 & 0.84 & 0.89 & 0.95 \\ 
& &   L2 $(n=300)$ & 0.49 & 0.54 & 0.59 & 0.64 & 0.69 & 0.74 & 0.79 & 0.84 & 0.89 & 0.95 \\ 
& &   Linf $(n=300)$ & 0.49 & 0.54 & 0.59 & 0.64 & 0.69 & 0.74 & 0.79 & 0.84 & 0.90 & 0.95 \\ \\
& $B_1=99$  &  L2 $(n=100)$ & 0.49 & 0.53 & 0.58 & 0.63 & 0.68 & 0.73 & 0.78 & 0.83 & 0.88 & 0.94 \\ 
& $B_2 = 99$ &   Linf $(n=100)$ & 0.48 & 0.53 & 0.58 & 0.63 & 0.68 & 0.73 & 0.78 & 0.84 & 0.89 & 0.94 \\ 
& &   L2 $(n=300)$ & 0.48 & 0.53 & 0.58 & 0.64 & 0.69 & 0.74 & 0.79 & 0.83 & 0.89 & 0.94 \\ 
& &   Linf $(n=300)$ & 0.49 & 0.54 & 0.59 & 0.63 & 0.68 & 0.73 & 0.78 & 0.84 & 0.89 & 0.94 \\ \\
Median & $B_1=399$  &  L2 $(n=100)$ & 0.33 & 0.41 & 0.50 & 0.59 & 0.65 & 0.70 & 0.75 & 0.82 & 0.87 & 0.92 \\ 
  (FM) & $B_2 = 399$ &  Linf $(n=100)$ & 0.36 & 0.43 & 0.51 & 0.59 & 0.65 & 0.70 & 0.75 & 0.81 & 0.86 & 0.91 \\ 
  & &   L2 $(n=300)$ & 0.40 & 0.45 & 0.52 & 0.61 & 0.66 & 0.70 & 0.75 & 0.80 & 0.86 & 0.91 \\ 
  & &   Linf $(n=300)$ & 0.40 & 0.45 & 0.51 & 0.59 & 0.64 & 0.68 & 0.72 & 0.77 & 0.83 & 0.88 \\ \\
  & $B_1=399$  &  L2 $(n=100)$ &  0.34 & 0.41 & 0.49 & 0.57 & 0.64 & 0.70 & 0.75 & 0.81 & 0.87 & 0.92 \\ 
  & $B_2 = 99$ &   Linf $(n=100)$ &  0.36 & 0.43 & 0.50 & 0.58 & 0.64 & 0.69 & 0.74 & 0.80 & 0.85 & 0.90 \\ 
  & &   L2 $(n=300)$   & 0.40 & 0.45 & 0.52 & 0.60 & 0.66 & 0.70 & 0.75 & 0.80 & 0.86 & 0.91 \\ 
  & &   Linf $(n=300)$  & 0.40 & 0.45 & 0.51 & 0.58 & 0.64 & 0.68 & 0.72 & 0.77 & 0.83 & 0.88 \\ \\
  & $B_1=99$  &  L2 $(n=100)$  & 0.36 & 0.44 & 0.53 & 0.62 & 0.68 & 0.72 & 0.77 & 0.83 & 0.89 & 0.93 \\ 
  & $B_2 = 399$&   Linf $(n=100)$  & 0.38 & 0.45 & 0.53 & 0.61 & 0.66 & 0.71 & 0.76 & 0.81 & 0.86 & 0.91 \\ 
  & &   L2 $(n=300)$  & 0.43 & 0.48 & 0.55 & 0.64 & 0.69 & 0.73 & 0.76 & 0.82 & 0.87 & 0.92 \\ 
  & &   Linf $(n=300)$  & 0.42 & 0.47 & 0.53 & 0.61 & 0.66 & 0.70 & 0.74 & 0.79 & 0.85 & 0.90 \\ \\
  & $B_1=99$  &  L2 $(n=100)$  & 0.33 & 0.40 & 0.49 & 0.57 & 0.64 & 0.69 & 0.75 & 0.81 & 0.87 & 0.92 \\ 
  & $B_2 = 99$ &   Linf $(n=100)$ & 0.36 & 0.42 & 0.50 & 0.57 & 0.64 & 0.69 & 0.74 & 0.80 & 0.85 & 0.90 \\ 
  & &   L2 $(n=300)$  & 0.39 & 0.45 & 0.51 & 0.58 & 0.65 & 0.69 & 0.74 & 0.79 & 0.85 & 0.91 \\ 
  & &   Linf $(n=300)$  & 0.40 & 0.45 & 0.51 & 0.57 & 0.63 & 0.67 & 0.71 & 0.76 & 0.82 & 0.87 \\ \\
Median & $B_1=399$  &  L2 $(n=100)$  & 0.59 & 0.64 & 0.68 & 0.73 & 0.77 & 0.82 & 0.86 & 0.89 & 0.93 & 0.97 \\ 
    ($\alpha$-radius) & $B_2 = 399$ &  Linf $(n=100)$ & 0.72 & 0.77 & 0.80 & 0.84 & 0.87 & 0.90 & 0.93 & 0.95 & 0.97 & 0.99 \\ 
    & &   L2 $(n=300)$  & 0.56 & 0.61 & 0.66 & 0.70 & 0.75 & 0.79 & 0.84 & 0.88 & 0.92 & 0.96 \\ 
    & &   Linf $(n=300)$ & 0.68 & 0.73 & 0.77 & 0.81 & 0.85 & 0.88 & 0.91 & 0.94 & 0.96 & 0.99 \\ \\
    & $B_1=399$  &  L2 $(n=100)$  & 0.58 & 0.63 & 0.68 & 0.72 & 0.76 & 0.81 & 0.85 & 0.89 & 0.92 & 0.96 \\ 
    & $B_2 = 99$ &   Linf $(n=100)$  & 0.71 & 0.76 & 0.80 & 0.83 & 0.87 & 0.90 & 0.92 & 0.95 & 0.97 & 0.99 \\ 
    & &   L2 $(n=300)$    & 0.56 & 0.60 & 0.65 & 0.70 & 0.74 & 0.79 & 0.83 & 0.87 & 0.91 & 0.95 \\ 
    & &   Linf $(n=300)$  & 0.67 & 0.72 & 0.76 & 0.80 & 0.84 & 0.87 & 0.91 & 0.93 & 0.96 & 0.98 \\ \\
    & $B_1=99$  &  L2 $(n=100)$   & 0.59 & 0.64 & 0.68 & 0.73 & 0.77 & 0.81 & 0.86 & 0.89 & 0.93 & 0.96 \\ 
    & $B_2 = 399$&   Linf $(n=100)$  & 0.72 & 0.76 & 0.80 & 0.84 & 0.87 & 0.90 & 0.93 & 0.95 & 0.97 & 0.99 \\ 
    & &   L2 $(n=300)$   & 0.56 & 0.61 & 0.66 & 0.70 & 0.75 & 0.79 & 0.84 & 0.88 & 0.92 & 0.96 \\ 
    & &   Linf $(n=300)$   & 0.68 & 0.72 & 0.76 & 0.81 & 0.84 & 0.88 & 0.91 & 0.94 & 0.97 & 0.99 \\ \\
    & $B_1=99$  &  L2 $(n=100)$   & 0.58 & 0.63 & 0.68 & 0.72 & 0.77 & 0.81 & 0.85 & 0.89 & 0.92 & 0.96 \\ 
    & $B_2 = 99$ &   Linf $(n=100)$  & 0.71 & 0.76 & 0.80 & 0.83 & 0.87 & 0.90 & 0.93 & 0.95 & 0.97 & 0.99 \\ 
    & &   L2 $(n=300)$   & 0.55 & 0.60 & 0.65 & 0.69 & 0.74 & 0.78 & 0.83 & 0.87 & 0.91 & 0.95 \\ 
    & &   Linf $(n=300)$ & 0.67 & 0.71 & 0.76 & 0.80 & 0.84 & 0.87 & 0.90 & 0.93 & 0.96 & 0.98 \\   \\
Trimmed  & $B_1=399$  &  L2 $(n=100)$   & 0.50 & 0.55 & 0.60 & 0.65 & 0.70 & 0.76 & 0.81 & 0.86 & 0.90 & 0.95 \\ 
    mean (FM) & $B_2 = 399$ &  Linf $(n=100)$  & 0.50 & 0.55 & 0.60 & 0.65 & 0.71 & 0.76 & 0.81 & 0.86 & 0.91 & 0.96 \\ 
    & &   L2 $(n=300)$  & 0.51 & 0.56 & 0.61 & 0.66 & 0.71 & 0.76 & 0.81 & 0.86 & 0.90 & 0.95 \\ 
    & &   Linf $(n=300)$  & 0.51 & 0.56 & 0.61 & 0.66 & 0.71 & 0.76 & 0.81 & 0.85 & 0.90 & 0.95 \\ \\
    & $B_1=399$  &  L2 $(n=100)$  & 0.50 & 0.55 & 0.59 & 0.65 & 0.70 & 0.75 & 0.80 & 0.85 & 0.90 & 0.95 \\ 
    & $B_2 = 99$ &   Linf $(n=100)$  & 0.49 & 0.55 & 0.59 & 0.65 & 0.70 & 0.75 & 0.80 & 0.85 & 0.90 & 0.95 \\ 
    & &   L2 $(n=300)$  & 0.50 & 0.55 & 0.60 & 0.65 & 0.70 & 0.75 & 0.80 & 0.85 & 0.90 & 0.95 \\ 
    & &   Linf $(n=300)$ & 0.50 & 0.55 & 0.60 & 0.65 & 0.70 & 0.75 & 0.80 & 0.85 & 0.90 & 0.94 \\ \\
    & $B_1=99$  &  L2 $(n=100)$   & 0.50 & 0.55 & 0.60 & 0.65 & 0.70 & 0.75 & 0.80 & 0.85 & 0.90 & 0.95 \\ 
    & $B_2 = 399$&   Linf $(n=100)$  & 0.50 & 0.55 & 0.60 & 0.65 & 0.70 & 0.75 & 0.81 & 0.86 & 0.91 & 0.95 \\ 
    & &   L2 $(n=300)$ & 0.50 & 0.55 & 0.60 & 0.65 & 0.71 & 0.75 & 0.81 & 0.85 & 0.90 & 0.95 \\ 
    & &   Linf $(n=300)$   & 0.51 & 0.56 & 0.61 & 0.66 & 0.71 & 0.76 & 0.80 & 0.85 & 0.90 & 0.95 \\ \\
    & $B_1=99$  &  L2 $(n=100)$   & 0.49 & 0.54 & 0.59 & 0.64 & 0.70 & 0.75 & 0.80 & 0.85 & 0.89 & 0.94 \\ 
    & $B_2 = 99$ &   Linf $(n=100)$ & 0.49 & 0.54 & 0.59 & 0.64 & 0.69 & 0.75 & 0.80 & 0.85 & 0.90 & 0.95 \\ 
    & &   L2 $(n=300)$ & 0.49 & 0.54 & 0.59 & 0.65 & 0.70 & 0.75 & 0.80 & 0.85 & 0.90 & 0.94 \\ 
    & &   Linf $(n=300)$   & 0.50 & 0.55 & 0.60 & 0.65 & 0.70 & 0.75 & 0.80 & 0.85 & 0.90 & 0.94 \\    \\
Trimmed &     $B_1=399$  &  L2 $(n=100)$  & 0.51 & 0.56 & 0.61 & 0.66 & 0.71 & 0.76 & 0.81 & 0.86 & 0.91 & 0.95 \\
    mean  & $B_2 = 399$ &  Linf $(n=100)$  & 0.51 & 0.56 & 0.61 & 0.66 & 0.71 & 0.77 & 0.82 & 0.87 & 0.91 & 0.96 \\
($\alpha$-radius)    & &   L2 $(n=300)$  & 0.51 & 0.56 & 0.61 & 0.66 & 0.71 & 0.76 & 0.81 & 0.86 & 0.90 & 0.95 \\
    & &   Linf $(n=300)$  & 0.51 & 0.56 & 0.61 & 0.66 & 0.71 & 0.76 & 0.81 & 0.86 & 0.91 & 0.95  \\\\
       & $B_1=399$  &  L2 $(n=100)$  & 0.50 & 0.55 & 0.60 & 0.65 & 0.70 & 0.75 & 0.80 & 0.85 & 0.90 & 0.95 \\
           & $B_2 = 99$ &   Linf $(n=100)$  & 0.50 & 0.55 & 0.60 & 0.65 & 0.71 & 0.76 & 0.81 & 0.86 & 0.90 & 0.95 \\
      & &   L2 $(n=300)$  & 0.50 & 0.55 & 0.60 & 0.65 & 0.70 & 0.75 & 0.80 & 0.85 & 0.90 & 0.94 \\
        & &   Linf $(n=300)$ & 0.50 & 0.55 & 0.60 & 0.65 & 0.70 & 0.75 & 0.80 & 0.85 & 0.90 & 0.95  \\\\
        & $B_1=99$  &  L2 $(n=100)$  & 0.50 & 0.55 & 0.61 & 0.66 & 0.71 & 0.76 & 0.81 & 0.86 & 0.90 & 0.95  \\
       & $B_2 = 399$&   Linf $(n=100)$   & 0.50 & 0.56 & 0.61 & 0.66 & 0.71 & 0.76 & 0.82 & 0.86 & 0.91 & 0.96 \\
             & &   L2 $(n=300)$ & 0.50 & 0.55 & 0.61 & 0.66 & 0.71 & 0.76 & 0.80 & 0.85 & 0.90 & 0.95 \\
     & &   Linf $(n=300)$  & 0.51 & 0.56 & 0.61 & 0.66 & 0.71 & 0.76 & 0.81 & 0.86 & 0.91 & 0.95  \\\\
    & $B_1=99$  &  L2 $(n=100)$   & 0.50 & 0.55 & 0.60 & 0.65 & 0.70 & 0.75 & 0.80 & 0.85 & 0.90 & 0.94 \\
     & $B_2 = 99$ &   Linf $(n=100)$  & 0.50 & 0.55 & 0.61 & 0.66 & 0.71 & 0.76 & 0.80 & 0.86 & 0.90 & 0.95 \\
    & &   L2 $(n=300)$ & 0.49 & 0.55 & 0.60 & 0.65 & 0.70 & 0.75 & 0.80 & 0.84 & 0.89 & 0.94  \\
    & &   Linf $(n=300)$  & 0.50 & 0.55 & 0.60 & 0.65 & 0.70 & 0.75 & 0.80 & 0.85 & 0.90 & 0.94   \\\\                                                
Variance & $B_1=399$  &  L2 $(n=100)$ & 0.46 & 0.51 & 0.56 & 0.61 & 0.67 & 0.72 & 0.77 & 0.83 & 0.88 & 0.93 \\ 
  & $B_2 = 399$ &  Linf $(n=100)$  & 0.48 & 0.53 & 0.58 & 0.64 & 0.69 & 0.75 & 0.80 & 0.86 & 0.91 & 0.96 \\ 
      & &   L2 $(n=300)$   & 0.48 & 0.53 & 0.58 & 0.64 & 0.69 & 0.74 & 0.79 & 0.84 & 0.89 & 0.94 \\ 
      & &   Linf $(n=300)$  & 0.49 & 0.54 & 0.59 & 0.64 & 0.70 & 0.75 & 0.80 & 0.86 & 0.91 & 0.96 \\ \\
      & $B_1=399$  &  L2 $(n=100)$   & 0.45 & 0.50 & 0.55 & 0.61 & 0.66 & 0.71 & 0.76 & 0.82 & 0.87 & 0.92 \\ 
      & $B_2 = 99$ &   Linf $(n=100)$ & 0.47 & 0.52 & 0.58 & 0.63 & 0.68 & 0.74 & 0.79 & 0.85 & 0.90 & 0.95 \\ 
      & &   L2 $(n=300)$ & 0.48 & 0.53 & 0.58 & 0.63 & 0.68 & 0.73 & 0.78 & 0.83 & 0.88 & 0.94 \\ 
      & &   Linf $(n=300)$ & 0.48 & 0.53 & 0.58 & 0.64 & 0.69 & 0.74 & 0.79 & 0.85 & 0.90 & 0.95 \\ \\
      & $B_1=99$  &  L2 $(n=100)$  & 0.46 & 0.51 & 0.56 & 0.61 & 0.67 & 0.72 & 0.77 & 0.83 & 0.88 & 0.93 \\ 
      & $B_2 = 399$&   Linf $(n=100)$  & 0.48 & 0.53 & 0.58 & 0.64 & 0.69 & 0.75 & 0.80 & 0.86 & 0.91 & 0.96 \\ 
      & &   L2 $(n=300)$  & 0.49 & 0.54 & 0.59 & 0.64 & 0.69 & 0.74 & 0.79 & 0.84 & 0.89 & 0.95 \\ 
      & &   Linf $(n=300)$  & 0.49 & 0.54 & 0.59 & 0.65 & 0.70 & 0.75 & 0.80 & 0.86 & 0.91 & 0.96 \\ \\
      & $B_1=99$  &  L2 $(n=100)$ & 0.45 & 0.50 & 0.55 & 0.60 & 0.66 & 0.71 & 0.76 & 0.82 & 0.87 & 0.93 \\ 
      & $B_2 = 99$ &   Linf $(n=100)$  & 0.47 & 0.52 & 0.58 & 0.63 & 0.69 & 0.74 & 0.79 & 0.85 & 0.90 & 0.95 \\ 
      & &   L2 $(n=300)$  & 0.48 & 0.53 & 0.58 & 0.63 & 0.68 & 0.73 & 0.78 & 0.83 & 0.89 & 0.94 \\ 
      & &   Linf $(n=300)$   & 0.48 & 0.53 & 0.59 & 0.64 & 0.69 & 0.74 & 0.80 & 0.85 & 0.90 & 0.95 \\     
\bottomrule
\end{longtable}
\end{center}
\end{small}

\end{appendix}

\newpage
\bibliographystyle{agsm}

\end{document}